\documentclass[12pt,a4paper]{article}
 \usepackage{jheppub}
 \usepackage{amsmath}
 \usepackage{amsfonts}
 \usepackage{mathtools}
 \usepackage{amssymb}
 \usepackage{caption}
 \usepackage{subcaption}
 \usepackage{slashed}
\usepackage{float}
\usepackage{blkarray}
\newcommand*{\ditto}{---\texttt{"}---}

\def\b{\bullet}
\def\al{\alpha}

\newcommand{\be} {\begin{equation}}
\newcommand{\ee} {\end{equation}}
\newcommand{\bea} {\begin{eqnarray}}
\newcommand{\eea} {\end{eqnarray}}
\newcommand{\ba} {\begin{array}}
\newcommand{\ea} {\end{array}}
\newcommand{\nn} {\nonumber}

\begin{document}
\title{Heterotic strings on $(K3\times T^2)/\mathbb{Z}_3$ and their dual  Calabi-Yau threefolds}
\author{Andreas Banlaki$^{\,a}$, Aradhita Chattopadhyaya$^{\,b}$, Abhiram Kidambi$^{\,a,d}$,\\{Thorsten Schimannek$^{\,c}$, Maria Schimpf$^{\,a}$}}
\affiliation{
	$^{a}$Institute for Theoretical Physics, TU Wien, A-1040 Vienna, Austria\\[.2em]
	$^{b}$Department of Mathematics, Trinity College Dublin, Dublin 2, Ireland\\[.2em]
	$^{c}$Faculty of Physics, University of Vienna, A-1090, Vienna, Austria\\[.2em]
    $^{d}$Stanford Institute for Theoretical Physics, Stanford University, Palo Alto, USA}
\emailAdd{banlaki@hep.itp.tuwien.ac.at, aradhita@maths.tcd.ie,\\abhiram.kidambi@tuwien.ac.at, mschimpf@hep.itp.tuwien.ac.at,\\thorsten.schimannek@univie.ac.at}
\abstract{
In this paper we study compactifications of the ${\cal N}=2$ heterotic $E_8\times E_8$ string on $(K3\times T^2)/\mathbb{Z}_3$ with various gauge backgrounds and calculate
the topological couplings in the effective supergravity action that arise from one-loop amplitudes.
We then identify candidates for dual type IIA compactifications on Calabi-Yau threefolds and compare the heterotic results with the corresponding topological string amplitudes.
We find that the dual Calabi-Yau geometries are $K3$ fibrations that are also genus one fibered with three-sections.
Moreover, we show that the intersection form on the polarization lattice of the $K3$ fibration has to be three times the intersection form on the Narain lattice $\Gamma^{1,1}$.

}
 UWThPh-2019-35\hfill
	\maketitle
	\flushbottom
\section{Introduction}
The four-dimensional $\mathcal{N}=2$ effective supergravity actions that arise from heterotic compactifications on $(K3\times T^2)/\mathbb{Z}_N$ and type II compactifications on Calabi-Yau threefolds contain the well-known supersymmetry protected couplings
\begin{align}
	S=\int F_g(y,\bar{y})\cdot F_+^{2g-2}R_+^2\,,
	\label{eqn:FRcoupling1}
\end{align}
where $F_+$ and $R_+$ respectively denote the self-dual parts of the graviphoton field strength and the Riemann tensor.
In particular, the coefficients $F_g(y,\bar{y})$ depend only on the vector moduli.
On the heterotic side, one of the vector moduli corresponds to the axio-dilaton and all terms~\eqref{eqn:FRcoupling1} that do not involve the dilaton
arise already at one-loop level.
The one-loop amplitudes have been evaluated for compactifications with various embeddings of the gauge connection on $K3\times T^2$ by~\cite{Harvey:1995fq,Marino:1998pg,Stieberger:1998yi}
and receive contributions only from BPS states that are encoded in the so-called new supersymmetric index~\cite{Cecotti:1992qh}.
Subsequently this analysis has been extended to compactifications on $(K3\times T^2)/\mathbb{Z}_N$~\cite{Chattopadhyaya:2017zul} and again the result is essentially encoded in the new supersymmetric index.
The indices in turn had been calculated for various compactifications with standard embedding of the spin connection into the gauge connection on $(K3\times T^2)/\mathbb{Z}_N$ and some non-standard embeddings when $N=2$~\cite{Chattopadhyaya:2016xpa}.
Calabi-Yau duals for some of the $N=2$ cases have been proposed in~\cite{Kachru:1997bz,Banlaki:2018pcc,Cota:2019cjx}.

In this paper we extend the calculation of the new supersymmetric index to compactifications on $(K3\times T^2)/\mathbb{Z}_3$ with non-standard embeddings and predict the geometric data of the dual Calabi-Yau manifold.
To this end we take an explicit realization of the orbifold limit of a $K3$ as $T^4/\mathbb{Z}_3$.
For two embeddings the effective theories are such that the gauge group on a generic point of the hypermultiplet moduli space is maximally broken to $U(1)^3$.
For those models one can hope to find a dual Calabi-Yau compactification with only three K\"ahler moduli and without a larger non-higgsable gauge group.
Indeed we will be able to identify many candidates.
At certain points in the moduli space there might be singularities as additional vector bosons become massless.
The strength of these singularities and the extraction of the Gopakumar-Vafa invariants are discussed in section \ref{hetgv}.

On the type IIA side one can identify $\lim_{y\rightarrow\infty}F_g(y,\bar{y})$ with the complex conjugate of the topological string
free energy at genus $g$ in the holomorphic limit~\cite{Bershadsky:1993ta,Bershadsky:1993cx,Antoniadis:1993ze,Marino:1998pg}.
Following arguments from~\cite{Cota:2019cjx} that we review in section~\ref{ssec:abriefreview}, one expects that for heterotic compactifications on $(K3\times T^2)/\mathbb{Z}_3$ the dual Calabi-Yau manifolds are genus one fibered with three sections.
Moreover, based on the prepotential of the effective supergravity action, one can argue that the Calabi-Yau should also exhibit a $K3$ fibration~\cite{Klemm:1995tj}.
We thus systematically construct all $K3$ fibered Calabi-Yau threefolds with $h^{1,1}=3$ that exhibit a genus one fibration with three-sections and are realized as hypersurfaces in toric ambient spaces.
We then apply the modular bootstrap that has been extended to genus one fibrations with multi-sections in~\cite{Cota:2019cjx} and obtain all-genus results for the topological string amplitudes.
This provides all order checks of the duality.

Let us note that on the heterotic side, the quotient acts as an order three symplectic automorphism on $K3$ together with a one-third shift along $T^2$.
Symplectic automorphisms of $K3$ manifolds have been classified and form subgroups of the Mathieu group $M_{23}$~\cite{MukaiK3}.
In particular, there is up to conjugacy only one automorphism of order three and this corresponds to the so-called $3A$ class.
It turns out that the automorphism group of a non-linear sigma model into $K3$ is actually larger~\cite{Gaberdiel:2011fg} and contains another element of order three.
However, we will restrict to the geometric case.

The paper is structured as follows. In section~\ref{sec:compZnew} we study heterotic compactifications on $(K3\times T^2)/\mathbb{Z}_3$ where the $K3$ is realized in the orbifold limit as $T^4/\mathbb{Z}_3$.
We first calculate the twisted elliptic genus of $K3$ with order three twists by directly evaluating the corresponding trace.
This confirms earlier results from the literature that were based on indirect arguments and determines the new supersymmetric index for the standard embedding.
We then evaluate the new supersymmetric index for various non-standard embeddings.

In section~\ref{hetgv} we describe how the new supersymmetric index encodes the topological couplings and extract predictions for the Gopakumar-Vafa invariants of the dual Calabi-Yau compactification spaces.
We also discuss the singular behaviour of the generating functions at points where additional vector states become massless.

In section~\ref{sec:iiaduals} we review the relevant aspects of heterotic/type II duality and then systematically construct candidate Calabi-Yau duals with $h^{1,1}=3$ that in particular exhibit a $K3$
fibration and a compatible genus one fibration with three-sections.
We find generic expressions for the Gopakumar-Vafa invariants at genus zero that only depend on the Euler characterstic of the Calabi-Yau and the intersection form on the polarization lattice of the $K3$ fibration.
From the structure of the invariants we can see that for Calabi-Yau duals to heterotic compactifications on $(K3\times T^2)/\mathbb{Z}_3$,
the intersection form on the polarization lattice of the $K3$ fibration has to be three times the intersection form on the Narain lattice $\Gamma^{1,1}$.
Physically this condition can be related to the presence of T-duality on the heterotic side.
We will then apply the modular bootstrap to obtain all genus results for the Gopakumar-Vafa invariants of low degree with respect to the base of the $K3$ fiber.
We find matching invariants for both of the compactifications with non-standard embedding on the heterotic side that lead to a maximally higgsable gauge group
and for fifteen other geometries we expect that a heterotic dual exists.

The paper is supplemented by two Appendices.
In Appendix~\ref{app:conventions} we summarize the definitions of the various modular and Jacobi forms that appear throughout the discussion.
Finally, in Appendix~\ref{app:higgsing}, we discuss in some detail the higgsing of the gauge group for one of the models with non-standard embedding and give the necessary branching rules.

\vspace{.3cm}\noindent
{\bf Acknowledgements:} We thank Justin David, Rajesh Gopakumar, Timm Wrase, Harald Skarke, Cesar Fierro Cota and Albrecht Klemm for helpful discussions.
The work of T.S. is supported by the Austrian Science Fund (FWF): P30904-N27.
A.C. thanks the Centre for High Energy Physics at the Indian Institute of Science where a large portion of her work was done.
A.C. is supported by the IRC Laurette fellowship.
The work of A.B., A.C., A.K. and M.S. is supported by the OeAD `Scientific \& Technological Cooperation with India' grant Project IN 27/2018 titled `Mathieu moonshine and $\mathcal N = 2$ Heterotic - Type II string duality'. The work of A.K. was further supported by DKPI (Vienna) and the Austrian Marshall Fellowship.

\section{The new supersymmetric index for CHL orbifolds}
\label{sec:compZnew}
In this section we compute the massless spectrum as well as the new supersymmetric index for compactifications of the heterotic $E_8\times E_8$ string on $(K3\times T^2)/\mathbb{Z}_3$.
The generator $g'$ of $\mathbb{Z}_3$ acts via an order three shift along one of the circles of $T^2$ together with a symplectic automorphism of $K3$.
The latter corresponds to the unique order three subgroup of $M_{23} < M_{24}$ that is in the conjugacy class $3A$ of $M_{24}$.

We first compute the twisted elliptic genus of order $N=3$ orbifolds of $K3$.
The twisted elliptic genus is defined as the trace
\begin{equation}\label{teg}
F^{(r, s) }(\tau, z) =
\frac{1}{N}
Tr_{g'^r}((-1)^{F_L+\bar{F_R}} g'^s e^{2\pi i z F_L }q^{L_0}\bar{q}^{\bar{L}_0})\,,
\end{equation}
over the Ramond-Ramond sector of a non-linear sigma model into $K3$.
For $N=3$ the result can be expressed in terms of modular forms for $\Gamma_1(3)$ and reads
\begin{eqnarray}\label{tweg3}
F^{(0, 0) }(\tau, z)&=& \frac{8}{3}A(\tau,z),\qquad F^{(0, 1) }(\tau, z)= \frac{2}{3}A(\tau,z)-\frac{1}{2}B(\tau,z){\cal E}_3(\tau)\,,\\ \nn
&& F^{(r, rk) }(\tau, z)= \frac{2}{3}A(\tau,z)+\frac{1}{6}B(\tau,z){\cal E}_3\left(\frac{\tau+k}{3}\right),
\end{eqnarray}
where
\begin{equation}
A(\tau,z)=\frac{\theta_2 ^2(\tau,z)}{\theta_2 ^2(\tau,0)}+\frac{\theta_3 ^2(\tau,z)}{\theta_3 ^2(\tau,0)}+\frac{\theta_4 ^2(\tau,z)}{\theta_4 ^2(\tau,0)} \:,\: B(\tau,z)=\frac{\theta_1 ^2(\tau,z)}{\eta^6(\tau)}\,,
\end{equation}
while $r=1,2$ and $rk:=rk \mod 3$.
This result was essentially bootstrapped in~\cite{David:2006ji} and we reproduce it by directly evaluating the trace after taking a $T^4/\mathbb{Z}_3$ orbifold limit of the $K3$.
In particular, it encodes the new supersymmetric index for compactifications on $(K3\times T^2)/\mathbb{Z}_3$ with standard embedding, as predicted in~\cite{Chattopadhyaya:2017ews}.
The explicit calculation serves as a check that our construction of the orbifold is correct.
 
Next we move on to calculate the massless spectrum of the heterotic theory compactified on $(K3\times T^2)/\mathbb{Z}_3$ for various gauge backgrounds.
The realization of the $K3$ as $T^4/\mathbb{Z}_3$ allows for five inequivalent embeddings of $\mathbb{Z}_3$ into the gauge group which lead to an anomaly free theory~\cite{DIXON1986285,Stieberger:1998yi,Honecker:2006qz}.
They can be represented by \textit{shift vectors} $V=(\gamma,\tilde{\gamma})$ where $3\gamma$ and $3\tilde{\gamma}$ take values in the two $E_8$ lattices.
For the standard embedding, the difference between the number of hypermultiplets and vectormultiplets was computed in~\cite{Chattopadhyaya:2017zul}.
Among the four non-standard embeddings we find two models where the spectrum enables us to maximally higgs the gauge group.

Then we compute the new supersymmetric index of the heterotic theory compactified with standard embedding, where one of the $E_8$ groups remains unbroken.
The new supersymmetric index is given by
\begin{equation} \label{znew}
{\cal Z}_{{\rm new}}(\tau,\bar \tau)=\frac{1}{\eta^2(\tau)}{\rm Tr}_R\left((-1)^{F} F \, q^{L_0-c/24}\bar{q}^{\bar{L_0}-\bar{c}/24}\right)\,,
\end{equation}
where $q=e^{2\pi i \tau}$ and the trace is again performed over the Ramond sector of the internal CFT which has central charge $(c,\bar c)=(22,9)$.
The right moving worldsheet fermion number $F$ is given by the sum of the fermion number on $T^2$ and that of $K3$.
Our result matches that obtained by general arguments in~\cite{Chattopadhyaya:2016xpa} which acts as a further check to our orbifold construction.

We will then proceed to calculate the new supersymmetric index ${\cal Z}_{\rm new}$ for the four non-standard embeddings.
In each case of standard or non-standard embedding we can show that the difference between the number of hypermultiplets and vectormultiplets $N_h-N_v$
computed from the new supersymmetric index matches with the result obtained from explicitly counting the degrees of freedom.
 
\subsection{The twisted elliptic genus of $K3$}\label{twegsec}
We will now proceed to compute the twisted elliptic genus of the 3A orbifold of $K3$ where $K3$ is realized as  $T^4/\mathbb{Z}_3$.
Let us denote the coordinates on $T^4$ by $(z_1,z_2)\in\mathbb{C}^2/\mathbb{Z}^2$ with periodicity
\begin{align}
	z_i\sim z_i+n_1 e_1+n_2 e_2\,,\quad e_1=e^{\frac{2\pi i}{3}}\,,e_2=1\,,\quad(n_1,n_2)\in\mathbb{Z}^2\,,
\end{align}
i.e. the A- and B-cycles of the torus $e_1,e_2$ generate the root lattice of $SU(3)$.
Then the $\mathbb{Z}_3$ orbifold of $T^4$ which results in the orbifold limit of $K3$ can be implemented as
\begin{equation}
g^s:\,(z_1,\,z_2)\mapsto\left(e^{2\pi i s/3}z_1,\,e^{-2\pi i s/3}z_2\right)\,.
\label{t4z3}
\end{equation}

We also need to specify the action of the generator $g'$ of the $\mathbb{Z}_3$ that is used in the CHL quotient $(K3\times T^2)/\mathbb{Z}_3$.
We consider the $g'$ action as a one-third shift along the A-cycle and a two-third shift along the B-cycle,
\begin{equation}
	g':\,(z_1,\,z_2)\mapsto\left(z_1+\frac{1}{3}e_1+\frac{2}{3}e_2,\,z_2\right)\,,
\end{equation}
with
\begin{equation}
	g'^2:\,(z_1,\,z_2)\mapsto\left(z_1+\frac{2}{3}e_1+\frac{1}{3}e_2,\,z_2\right)\,.
\end{equation}
We first compute the twisted elliptic genus of the corresponding orbifold of $K3$ and show that it matches with the results obtained in~\cite{David:2006ji,David:2006ud,David:2006yn,Gaberdiel:2010ch,Eguchi:2010ej,Eguchi:2010fg,Cheng:2010pq}.
This serves as a check of our construction.

The twisted elliptic genus in the $(r,s)$-sector is given by the index
\begin{equation}\label{twgenus}
F^{(r, s) }(\tau, z) =
\frac{1}{9}\sum_{a,b = 0 }^2
Tr_{g^a,g'^r}\left((-1)^{F_L+\bar{F_R}}g^b g'^s e^{2\pi i z F_L }q^{L_0}\bar{q}^{\bar{L}_0}\right).
\end{equation}
 The overall factor of 1/9 comes from the projection of $g,g'$, both of which are ${\mathbb Z}_3$ orbifolds.
 The trace is taken over the Ramond-Ramond sector and $F_L, F_R$ are the left- and right-moving fermion numbers.
In defining the index we have suppressed the shifts $L_0 -c/24$ and $\bar L_0- \bar c/24$, where $(c,\bar c)=(6,6)$.
 
 Let us now define the trace 
 \begin{equation}
 {\cal F}( a, r; b,  s )  = \frac{1}{9}
 Tr_{g^a,g'^r}\left((-1)^{F_L+\bar{F_R}}g^b g'^s e^{2\pi i z F_L }q^{L_0}\bar{q}^{\bar{L}_0}\right)\,.
 \end{equation}
 To evaluate each of the above sectors in the twisted elliptic genus we will need the fixed points under the elements $g^ag^{\prime r}$, $a,r$ being the twists in $g$ and $g'$ respectively.
 We also require to know which elements preserve these fixed points.
 We summarize the result in  table \ref{fps}.
 
 \begin{table}[H] 
 	\renewcommand{\arraystretch}{0.5}
 	\begin{center}
 		\vspace{-0.1cm}
 		\begin{tabular}{|c|c|c|c|c|c|c|c|c|c|}
 			\hline
 			& Fixed points & $g'$ & $g'^2$ & $g$ & $g^2$ &  $g'g$ & $g'g^2$ & $g'^2g$ & $g'^2g^2$\\
 			\hline
 			& & & & & & & & & \\
 			$g$  & $(0,0)$, $(2/3,1/3)$, $(1/3,2/3)$   & $\times$ & $\times$ & \checkmark & \checkmark  &  $\times$ &  $\times$  &  $\times$ &  $\times$  \\
 			\hline
 			& & & & & & & & & \\
 			$g^2$  & $(0,0)$, $(2/3,1/3)$, $(1/3,2/3)$ & $\times$  & $\times$ & \checkmark & \checkmark  &  $\times$ &  $\times$  &  $\times$ &  $\times$ \\
 			 			\hline
 			& & & & & & & & & \\
			$gg'$  & $(2/3,0)$, $(0,2/3)$, $(1/3,1/3)$  & $\times$ & $\times$ & $\times$ & $\times$  &  \checkmark &  $\times$  &  $\times$ &  \checkmark \\
 				 			\hline
 				& & & & & & & & & \\
			$g^2g'$  & $(1/3,0)$, $(0,1/3)$, $(2/3,2/3)$ & $\times$  & $\times$ & $\times$ & $\times$  &  $\times$ & \checkmark  & \checkmark &  $\times$ \\
 					 			\hline
 					& & & & & & & & & \\
			$gg'^2$  & $(1/3,0)$, $(0,1/3)$, $(2/3,2/3)$   & $\times$ & $\times$ & $\times$ & $\times$  &  $\times$ & \checkmark  & \checkmark &  $\times$ \\
 										 			\hline
 										& & & & & & & & & \\
			$g^2g'^2$  & $(2/3,0)$, $(0,2/3)$, $(1/3,1/3)$ & $\times$  & $\times$ & $\times$ & $\times$  &  \checkmark &  $\times$  &  $\times$ &  \checkmark \\
 			\hline
 		\end{tabular}
 	\end{center}
 	\vspace{-0.5cm}
	 \caption{Rows list the  properties of points that are fixed under the action of the element $g^a{g'}^b$ in the first column.
	 A $\times$ denotes that the point moves under the action of the corresponding element ${g'}^ag^b$ in the top row, while $\checkmark$ indicates that the point remains fixed. 
 		Positions are given as multiples of the A- and B-cycle $e_1$ and $e_2$.}
 	\label{fps}
 	\renewcommand{\arraystretch}{0.5}
 \end{table}
 
Now we evaluate the different sectors of the twisted elliptic genus. The sector $(0, 0)$ comes with no twists in $g'$ or insertions of $g'$ and so the trace reduces to 
 \begin{equation}\label{tweg0}
 F^{(0,0)}(\tau, z) =\frac{1}{3}Z_{K3} (\tau, z) = \frac{8}{3} A(\tau, z)\,.
 \end{equation}
 where $Z_{K3}$ is the elliptic genus of $K3$.

 Let us now look into the untwisted sectors of the twisted elliptic genus. In the $(0, 1)$ and $(0,2)$ sectors we obtain from  table \ref{fps},
 that a single insertion of $g'$ does  not preserve any of the fixed points. 
 Hence we get,
 \begin{equation}
 {\cal F} ( a, 0; b, 1)= {\cal F} ( a, 0; b, 2)=0, \qquad {\rm for }\;  a = 1, 2\,.
 \end{equation}
 
 Evaluating the trace in the untwisted sector we see the contributions are 
 \begin{eqnarray}
 {\cal F} (0, 0; 0, s) &=& 0\,, \qquad  \\ \nonumber
 {\cal F} ( 0, 0; 1, 1)=  {\cal F} ( 0, 0; 1, 2)& = & 
 \frac{ \theta_1 ( z+ \frac{1}{3} , \tau) \theta_1( - z + \frac{1}{3} ) }
 { \theta_1^2 ( \frac{1}{3} , \tau ) }\,, \\ \nonumber
 {\cal F} ( 0, 0; 2, 1)=  {\cal F} ( 0, 0; 2, 2) &=&  \frac{ \theta_1 ( z+ \frac{2}{3} , \tau) \theta_1( - z + \frac{2}{3} ) }
 { \theta_1^2 ( \frac{2}{3} , \tau ) }\,.
 \end{eqnarray}
 The coefficients (numerical) in the traces come from the contribution of 
 fermionic zero modes. There are $9$ right moving fermionic zero modes when $g$ or $g^2$ are inserted in the trace.
 It can be easily shown by considering the $q$ expansions that
 \begin{eqnarray}
  {\cal F}( 0, 0; 1, 1)&+ &{\cal F} ( 0, 0; 2,1)={\cal F}( 0, 0; 1, 2)+ {\cal F} ( 0, 0; 2,2)\\ \nn
  &=&\frac{2}{3}A(\tau,z)-\frac{1}{2}B(\tau,z){\cal E}_3(\tau)
  =F_{3A}^{(0,1)}(\tau,z)=F_{3A}^{(0,2)}(\tau,z)\,,
 \end{eqnarray}
 as predicted in \cite{David:2006ji,David:2006ud,David:2006yn,Gaberdiel:2010ch,Eguchi:2010ej,Eguchi:2010fg}.
 
 The other sectors can be obtained from  $F^{(0,1)}$ through the modular transformation property of the twisted elliptic genus given by,
 \begin{equation} \label{fulltwist}
 F^{(r, s)} \left( \frac{a\tau + b}{ c\tau + d}, \frac{ z}{ c\tau +d}
 \right) = \exp\left(  2\pi i \frac{c z^2}{ c\tau +d}  \right) 
 F^{( cs + a r, ds + b r)} ( \tau,  z)\,,
 \end{equation}
 where  $a, b, c, d \in \mathbb{Z}, \; ad -bc =1.$
 However, with the knowledge of the fixed points we can directly compute the twisted sectors and re-evaluate the results of the twisted elliptic genus in each sector.
 
 In the sectors twisted by $g'$ without any insertion of $g'$ we have the components of the twisted elliptic genus
 given by
 \be \nn
F^{(1,0)}=\sum_{a,b =0}^2 {\cal F}(a,1,b,0)\,.
 \ee
 Now ${\cal F}(0,1,1,0)=0={\cal F}(0,1,2,0)$ and also ${\cal F}(1,1,1,0)=0={\cal F}(2,1,1,0)$ due to the absence of any fixed point in the twisted sector (table \ref{fps}).
 The aame argument holds for ${\cal F}(1,1,2,0)=0={\cal F}(2,1,2,0)$. Also ${\cal F}(0,1,0,0)=0$ due to the presence of zero modes of the right moving fermions.
 Therefore the only contributions are
  \be \nn
 F^{(1,0)}={\cal F}(1,1,0,0)+{\cal F}(2,1,0,0)\,.
 \ee

 Computing the trace we get
\begin{eqnarray}
{\cal F} (1,1,1,1)& = & 
\frac{ \theta_1 ( z+ \frac{\tau}{3} , \tau) \theta_1( - z + \frac{\tau}{3} ) }
{ \theta_1^2 ( \frac{\tau}{3} , \tau ) }\,, \\ \nonumber
{\cal F} (2,1,0,0) &=&  \frac{ \theta_1 ( z+ \frac{2\tau}{3} , \tau) \theta_1( - z + \frac{2\tau}{3} ) }
{ \theta_1^2 ( \frac{2\tau}{3} , \tau ) }\,,\\ \nn
 F^{(1,0) } &=&\frac{2}{3}A(\tau,z)+\frac{1}{6}B(\tau,z){\cal E}_3\left(\frac{\tau}{3}\right)\,.
\end{eqnarray}
If we insert a $g'$ in the trace we have the non-vanishing components in the twisted elliptic genus given by
\begin{eqnarray}
{\cal F} (1,1,1,1)& = & 
\frac{ \theta_1 ( z+ \frac{\tau+1}{3} , \tau) \theta_1( - z + \frac{\tau+1}{3} ) }
{ \theta_1^2 ( \frac{\tau+1}{3} , \tau ) }\,, \\ \nonumber
{\cal F} (2,1,2,1) &=&  \frac{ \theta_1 ( z+ \frac{2\tau+2}{3} , \tau) \theta_1( - z + \frac{2\tau+2}{3} ) }
{ \theta_1^2 ( \frac{2\tau+2}{3} , \tau ) }\,,\\ \nn
F^{(1,1)}  &=&\frac{2}{3}A(\tau,z)+\frac{1}{6}B(\tau,z){\cal E}_3\left(\frac{\tau+1}{3}\right)\,.
	\end{eqnarray}
The rest of the sectors vanish as all the fixed points change position.
Similarly for the insertion of $g'^2$ we have
\begin{eqnarray}
{\cal F} (1,1,1,1)& = & 
\frac{ \theta_1 ( z+ \frac{\tau+2}{3} , \tau) \theta_1( - z + \frac{\tau+2}{3} ) }
{ \theta_1^2 ( \frac{\tau+2}{3} , \tau ) }\,, \\ \nonumber
{\cal F} (2,1,2,1) &=&  \frac{ \theta_1 ( z+ \frac{2\tau+1}{3} , \tau) \theta_1( - z + \frac{2\tau+1}{3} ) }
{ \theta_1^2 ( \frac{2\tau+1}{3} , \tau ) }\,,\\ \nn
F^{(1,2) } &=&\frac{2}{3}A(\tau,z)+\frac{1}{6}B(\tau,z){\cal E}_3\left(\frac{\tau+2}{3}\right)\,.
	\end{eqnarray}
 A very similar argument holds for the $g'^2$ twisted sector and the results can be summarized as:
 \begin{eqnarray}
  F^{(2,0)} &=&\frac{2}{3}A(\tau,z)+\frac{1}{6}B(\tau,z){\cal E}_3\left(\frac{\tau}{3}\right)\,,\\ \nn
  F^{(2,1)} &=&\frac{2}{3}A(\tau,z)+\frac{1}{6}B(\tau,z){\cal E}_3\left(\frac{\tau+2}{3}\right)\,,\\ \nn
  F^{(2,2)}  &=&\frac{2}{3}A(\tau,z)+\frac{1}{6}B(\tau,z){\cal E}_3\left(\frac{\tau+1}{3}\right)\,.
 \end{eqnarray}
 We note that these results were obtained previously in \cite{David:2006ji,David:2006ud,David:2006yn,Gaberdiel:2010ch,Eguchi:2010ej,Eguchi:2010fg},
 however, the authors are not aware of any paper that explicitly evaluates the traces.
 
 \subsection{Calculating the massless spectrum}
In this section we derive the massless spectrum of the heterotic string theory compactified on $(K3\times T^2)/\mathbb{Z}_N$.
To this end we consider again an orbifold limit of the $K3$ such that the compactification space is $\left[(T^4/\mathbb{Z}_3)\times T^2\right]/\mathbb{Z}_3$.
The orbifold action on the six toroidal coordinates can be given by
 \be \label{eqn:tor_act}
 g:\,(z_1,\,z_2,\,z_3)\mapsto(e^{2\pi i/3}z_1,\,e^{-2\pi i/3}z_2,\,z_3)\,,
 \ee
 where $z_3$ is a coordinate on $T^2$ and $z_1,z_2$ parametrize the $T^4$.
 Again, the cycles of the $T^4$ form an $SU(3)$ lattice in $\mathbb{C}^2$ and we denote the A- and B-cycle by
 \begin{align}
	 e_1=e^{\frac{2\pi i}{3}}\,,\quad e_2=1\,.
 \end{align}
 We can assume that the corresponding lattice for the $T^2$ is generated by $1$ and $i$.
 The action of the CHL quotient is then
 \be \label{eqn:chl_act}
 g^\prime:\,(z_1,\,z_2,\,z_3)\mapsto\left(z_1+\frac{1}{3}e_1+\frac{2}{3}e_2,\,z_2,\,z_3+\frac{1}{3}\right)\,.
 \ee
 The spectrum of $K3$ realized as $T^4/\mathbb{Z}_3$ with the standard shift was first studied in~\cite{Walton:1987bu} and we will follow the discussion of~\cite{Aldazabal:1995yw,Walton:1987bu}.
 Note that the orbifold action $g'$ does not produce any fixed points and so it preserves ${\cal N}=2$ supersymmetry. 
 
 The massless spectrum organizes into four-dimensional ${\cal N}=2$ supergravity multiplets with $N_v$ vectormultiplets and $N_h$ hypermultiplets in the theory.
The massless states in the $g$ twisted sector are determined by setting left and right masses to zero
 \begin{align}
	 \begin{split}
 m_L ^2&=N_L+\frac{1}{2}(P+nV)^2+E_n-1  =0\,, \\
 m_R ^2&=N_R+\frac{1}{2}(r+nv)^2+E_n-\frac{1}{2} =0\,.
\label{lrmass}
	 \end{split}
 \end{align}
 Here $P$ is an $E_8\times E_8$ lattice vector and of the form
 \begin{equation}
 P = (P_{E_8} ; P_{E_8'})\,,
 \end{equation}
while $E_n$ is the shift in zero point energy on the ground state due to the twisting by $g^n$.
It is given by 
 \begin {equation}\label{zpt}
 E_n = \frac{1}{9} n (\nu - n ) ,
 \end {equation}
 where $\nu=3$ for the $T^4/\mathbb{Z}_3$ orbifold and $ n = 0, 1, 2$. 
Moreover, $r$ is an $SO(8)$ weight vector with 
 \begin{equation}
 \sum_{i=1}^4r_i =\;  {\rm odd}\,,
 \end{equation}
and the $4$ dimensional vector $v$ with
 \begin{equation}\label{defv}
 v = \frac{1}{3} (0,0,1,-1)\,,
 \end{equation}
 encodes the action that we used to define the $T^4/\mathbb{Z}_3$ orbifold.
 The value of $V$ encodes the embedding of the $\mathbb{Z}_3$ into the gauge group.
 It takes values in the two $E_8$ lattices scaled by one third.
For standard embedding we have the shift given by
 \begin{equation} \label{shift}
 V=\frac{1}{3}(1,-1,0^6;0^8)\,.
 \end{equation}
 The degeneracy of the massless states for $T^4/\mathbb{Z}_3$ model can be obtained from \cite{Aldazabal:1995yw}
 \begin{eqnarray}\label{unmod}
 D(n) &=&\frac{1}{3}\sum_{m=0}^{2}\chi(n,m)\Delta(n,m), \\ \nn
 \Delta(n,m)&=&\exp\left\{2\pi i[(r+nv)mv-(P+nV)mV+\frac{1}{2}mn(V^2-v^2)+m\rho]\right\}.
 \end{eqnarray}
 Here $\chi(n, m)$ are the number of fixed points in the $g^n$ twisted sector invariant under the action of $g^m$ and $\rho$ is the phase by which oscillators are rotated under the $\mathbb{Z}_3$ action on the $T^4$.
 
 In the case of order three CHL orbifolds this formula is modified to
 \begin{equation}\label{dgen3}
 D(n;g' ) = \frac{1}{3}\sum_{m=0}^{2}\frac{1}{3}\left[ \chi(n,m) + \chi^{(g')}(n,m)+ \chi^{(g'^2)} (n, m) \right]\Delta(n,m)\,,
 \end{equation}
 where $\chi^{(g'^s)}(n,m)$ is the number of fixed points in the $g^n $ twisted sector invariant under the action of $g^m g'^s$.
 Essentially we implement the projection under the  $g'$ action. In the untwisted sector 
 \begin{equation}\label{untwg}
 \chi^{(g'^s)} ( 0, m )  =  \chi(0,m) =1,\; {\rm for}\; s=1,2\,,
 \end{equation} 
 and the phases in (\ref{dgen3}) implement how the spectrum is projected under $g^m$.
 In the twisted sectors we can use again table~\ref{fps} to obtain
 \begin{eqnarray} \label{modded}
 &&\chi^{(g')}( 1 ,  m ) = \chi^{(g'^2)} (1, m ) = 0\,.
 \end{eqnarray}
 The sector $n=2$ gives the anti-particles of the $n=1$ sector.
 The untwisted sector, which counts the gravity multiplet and the number of vectors, remains the same as obtained in~\cite{Aldazabal:1995yw}.
 The result for the twisted sectors are just one third of that in~\cite{Aldazabal:1995yw}.
 
 In the following table~\ref{t4z3nh} we summarize the untwisted and twisted hypermultiplet contents in the various embeddings of $K3\times T^2$ where $K3$ is considered at the $T^4/\mathbb{Z}_3$ orbifold limit.
  \begin{table}[H]
 	\renewcommand{\arraystretch}{0.5}
 	\begin{center}
 		\vspace{0.5cm}
 		\begin{tabular}{|c|c|c|}
 			\hline
 			& & \\
 			Group, shift & Untwisted sector & Twisted sector \\ \hline
 			& & \\
 			$E_7\times U(1)\times E_8$& $(56;1)+2(1;1)$ & $9(56;1)$ \\
 			$\frac{1}{3}(1,-1,0^6;0^8)$ & $+(1;1)$ & + $45(1,1)$ \\
 			& & $18(1,1)$ \\ \hline
 			& & \\
 			$SU(9)\times E_8$ & $(84,1) + 2 (1,1)$ & $9 (36,1) + 18 (9,1)$\\
 		    $\frac 1 3 (2,1 ^4 , 0^3;0^8)$ & & \\
 		    & & \\\hline
 			& & \\
 			$SO(14)^2\times U(1)^2$ & $(14,1)+(1,14)$ & $9(14,1)$ \\
 			$\frac{1}{3}(2,0^7;2,0^7)$ & $+(64,1)+(1,64)$ & $+9(1,14)$ \\
 			& $+2(1,1)$ & $+18(1,1)$ \\ \hline
 			& & \\
 			$E_6\times SU(3)\times E_7\times U(1)$ & $(27,3,1)$ & $9(27,1,1)$ \\
 			$\frac{1}{3}(2,1^2,0^5;1,-1,0^6)$ & $+(1,1,56)$ & $9(1,3,1)$ \\
 			& $+2(1)+(1)$ & $+18(1,3,1)$ \\ \hline
 			& & \\
 			$SU(9)\times E_6\times SU(3)$ & $(84,1,1)+(1,27,3)$ & $9(9,1,3)$\\
 			$\frac{1}{3}(2,1^2,0^5;2,1^4,0^3)$ & $2(1)$ & \\ \hline
 		\end{tabular}
 	\end{center}
 	\vspace{-0.5cm}
 	\caption{Hypermultiplet spectrum for different embeddings with $K3$ as $T^4/\mathbb{Z}_3$.  We have not kept track of various U(1) charges.}\label{t4z3nh}
 	\renewcommand{\arraystretch}{0.5}
 \end{table}
 
 In the table~\ref{3anh} we show how the twisted sectors get modified under the order three orbifold action.
 \begin{table}[H]
 	\renewcommand{\arraystretch}{0.5}
 	\begin{center}
 		\vspace{0.5cm}
 		\begin{tabular}{|c|c|c|c|}
 			\hline
 			& & &\\
 			Group, shift & Untwisted sector & Twisted sector & $N_h-N_v$ \\
 			 \hline
 			&  &  & \\
 			$E_7\times U(1)\times E_8$ & $(56;1)+2(1;1)$ &  $3(56;1)$ & -134\\
 			$\frac{1}{3}(1,-1,0^6;0^8)$ & $+(1;1)$ &  +$15(1,1)$ &\\
 			&  & +$6(1,1)$ & \\ \hline
 			& & & \\
 			$SU(9)\times E_8$ & $(84,1) + 2 (1,1)$ & $3 (36,1) + 6 (9,1)$ & -80\\
 			$\frac 1 3 (2,1^4,0^3;0^8)$ & & & \\
 			& & & \\\hline
 			& & & \\
 			$SO(14)^2\times U(1)^2$ & $(14,1)+(1,14)$ &  $3(14,1)$  & 64\\
 			$\frac{1}{3}(2,0^7;2,0^7)$ & $+(64,1)+(1,64)$ &  $+3(1,14)$  &\\
 			& $+2(1,1)$ & $+6(1,1)$ & \\ \hline
 			& &  &\\
 			$E_6\times SU(3)\times E_7\times U(1)$ & $(27,3,1)$  & $3(27,1,1)$ & 28\\
			$\frac{1}{3}(2,1^2,0^5;1,-1,0^6)$ & $+(1,1,56)$  & $+3(1,3,1)$ & \\
 			& $+2(1)+(1)$  & $+6(1,3,1)$ & \\ \hline
 			& & & \\
			$SU(9)\times E_6\times SU(3)$ & (84,1,1)+(1,27,3)  & 3(9,1,3) & 82\\
			$\frac{1}{3}(2,1^2,0^5;2,1^4,0)$ & 2(1) &  &  \\ \hline
 		\end{tabular}
 	\end{center}
 	\vspace{-0.5cm}
 	\caption{Hypermultiplet spectrum for different embeddings with 3A orbifold of $K3$ where $K3$ is realized as $T^4/\mathbb{Z}_3$. We have not kept track of various U(1) charges.}\label{3anh}
 	\renewcommand{\arraystretch}{0.5}
 \end{table}

 The sectors twisted by $g'$ do not have massless modes as the winding numbers become one-third of integers. The untwisted sector contains the ${\cal N}=2$ gravity multiplet and ${\cal N}=2$ vector multiplets along with the hypers as obtained in \cite{Honecker:2006qz}. The twisted sectors only contain the hypers. The hypermultiplet contents are summarized in table \ref{t4z3nh}, \ref{3anh}. 
 Evaluating the difference between  $N_h$ and $ N_v$ we have for the standard embedding
 \be
 N_h-N_v=-134.
 \ee
 This was predicted in \cite{Chattopadhyaya:2016xpa}.
 Note that when counting the number of vector multiplets $N_v$ we only consider those that originate from the $E_8\times E_8$ group and not the four generic vectors that arise from the $T^2$.
 When no orbifold acts on $K3$ the theory can be lifted to six dimensions. The value of $N_h-N_v$ then remains 244 for any realization of $K3$ and for any embeddings because of anomaly cancellation in six dimension~\cite{Green:1984bx}. This is different for $N=2$ or $N=3$ orbifold models.

 By explicitly analysing the possible higgs chains for the gauge groups and spectra in table~\ref{3anh} we found that
 for two of the models the gauge group can be completely broken.
 These are the non-standard embeddings with $N_h-N_v=64$ and $N_h-N_v=82$.
 For the model with $N_h-N_v=64$ we provide the details of the higgs chain in Appendix~\ref{3anh}.
In section~\eqref{sec:iiaduals} we will construct Calabi-Yau geometries for which the enumerative invariants exactly  match the predictions
from the heterotic calculation of the gauge gravitational couplings for both of the models.
We also find many more Calabi-Yau manifolds for which we expect that a heterotic dual compactification
on $(K3\times T^2)/\mathbb{Z}_3$ exists but where the embedding can not be realized with our construction.
 
 \subsection{The new supersymmetric index}
 In this section we shall evaluate the new supersymmetric index of the heterotic string compactified on $K3\times  T^2$ with an order 3 CHL action on $K3\times T^2$ with various embeddings of $E_8\times E_8$. We shall explicitly compute the results of standard embeddings and summarize the result of non-standard ones.
 We consider $K3$ at the orbifold limit of $T^4/\mathbb{Z}_3$ where the $\mathbb{Z}_3$ action on $T^4$ is given in (\ref{eqn:tor_act}).
 The CHL action is given by (\ref{eqn:chl_act}).
 The new supersymmetric index is defined as
 \begin{equation} 
 {\cal Z}_{{\rm new}}(\tau,\bar\tau)=\frac{1}{\eta^2(\tau)}{\rm Tr}_R((-1)^{F} F q^{L_0-c/24}\bar{q}^{\bar{L_0}-\bar{c}/24}).
 \end{equation} 
 Under the standard embedding we have the shift action given by equation (\ref{shift}).
The above trace for standard embedding splits into the following sectors:
 \begin{align}
 {\cal Z}_{\rm new}(\tau, \bar \tau)  = -\frac{1}{2\eta^{20}(\tau)}\sum_{a,b=0}^2 \sum_{r,s=0}^2 e^{-\frac{2\pi iab}{9}} 
	 Z_{E_8}^{a,b}(\tau) \times E_4(\tau) 
 \times \frac{1}{9} F(a,r,b,s,\tau) \Gamma^{(r,s)}_{2,2} (q,\bar q)\,,\label{z3std}
 \end{align}
where only $\Gamma^{(r,s)}_{2,2} (q,\bar q)$ , defined in (\ref{gamma22}), has a non-holomorphic dependence which encodes the winding and momenta modes contribution of the $T^2$ bosons. The factor of $E_4$ is obtained from the untouched $E_8$ lattice.
We can write the trace over the directions of $T^4$ as
 \begin{equation} \label{fabrs}
 F(a, r, b, s; \tau) = {\rm Tr}_{g^a \, g^{\prime r}} 
 \left( g^b g^{\prime s} e^{i \pi F^{T^4}} q^{L_0} \bar q ^{\bar L_0}  \right)\,,
 \end{equation}
 and one can work out the explicit expression
 \begin{eqnarray} \label{fabrshet}
 F ( a, r, b, s; \tau)  =
 k^{a,r,b,s}\eta^2(\tau)q^{\frac{-a^2}{9}}\frac{1}{\theta_1 ^2(\frac{a\tau+b}{3}, \tau)}\,.
 \end{eqnarray}
For standard embeddings the coefficients $k^{a,r,b,s}$ for different values of $(r, s)$ are given by 
 the following matrices:
 \begin{align}
	 \begin{split}
		 k^{a,0,b,1}=&k^{a,0,b,2}=9\left(\begin{matrix} 0 & 1 & 1 \\ 0 & 0 & 0\\ 0 & 0 & 0
 \end{matrix} \right)\,,\quad k^{a,1,b,0}=k^{a,2,b,0}=9\left(\begin{matrix} 0 & 0 & 0 \\ 1 & 0 & 0\\ 1 & 0 & 0
 \end{matrix} \right)\,,\\
		 k^{a,1,b,1}=&k^{a,2,b,2}=9\left(\begin{matrix} 0 & 0 & 0 \\ 0 & 1 & 0\\ 0 & 0 & 1
 \end{matrix} \right)\,,\quad k^{a,2,b,1}=k^{a,1,b,2}=9\left(\begin{matrix} 0 & 0 & 0 \\ 0 & 0 & 1\\ 0 & 1 & 0
 \end{matrix} \right)\,,\\
		 k^{a,0,b,0}=&9\left(\begin{matrix} 0 & 1 & 1 \\ 1 & 1 & 1\\ 1 & 1 & 1
 \end{matrix} \right)\,.
	 \end{split}
 \end{align}
 Here the rows and columns are denoted by $a$ and $b$ respectively. 
 The result for the coefficients $k^{a,r,b,s}$ comes from counting the fixed points as for the case of twisted elliptic genus in table \ref{fps}.
 
 However, if the embedding is non-standard the new supersymmetric index is given by
 \begin{align}
	 {\cal Z}_{\rm new}(\tau, \bar \tau)  = -\frac{1}{2\eta^{20}(\tau)}\sum_{a,b,r,s=0}^2 e^{-2\pi i\frac{ab}{\nu^2}\Gamma^2} 
	 Z_{E_8}^{a,b}(\tau)  Z_{E_8'}^{a,b}(\tau) \frac{1}{\nu^2} F(a,r,b,s,\tau) \Gamma^{(r,s)}_{2,2}(\tau,\bar{\tau})\,,
 \label{z3ns}
 \end{align}
where the partition functions over the shifted $E_8$ lattices are 
 \begin{align}
	 \begin{split}
 Z_{E_8}^{a,b}(\tau)=&\sum_{\al,\beta=0}^{1}
 e^{-i\pi\beta a\sum_{I=1}^8\gamma^{I}}
 \prod_{I=1}^8\theta\left[
 \begin{smallmatrix}\al+2 a\gamma^I \\ \beta+2b\gamma^I\end{smallmatrix}\right],\\ 
 Z_{E_8'}^{a,b}(\tau)=&\sum_{\al,\beta=0}^{1}e^{-i\pi\beta a
 	\sum_{I=1}^8\tilde{\gamma}^{I}}\prod_{I=1}^8\theta\left[
 \begin{smallmatrix}\al+2a \tilde{\gamma}^I \\ \beta+2b\tilde{\gamma}^I
 \end{smallmatrix}\right],
	 \end{split}
 \end{align}
 with $\gamma, \tilde\gamma$ being the shifts in the $E_8, E_8'$ lattices, while $\nu=3$ is the order of the orbifold on $T^4$. 
 Moreover, $\Gamma^2=\nu^2(\gamma^2+\tilde{\gamma^2})$ is the sum of the squares of all the shifts in the two $E_8$ lattices.
 Under non-standard embeddings we have
 \begin{align}
	 \begin{split}
 k^{a,0,b,1}=& k^{a,0,b,2}=9\left(\begin{matrix}
 	0 & 1 & 1 \\
 	0 & 0 & 0\\
 	0 & 0 & 0
 \end{matrix}\right)\,,\quad k^{a,1,b,0}= k^{a,2,b,0}=9\left(\begin{matrix}
 0 & 0 & 0 \\
 1 & 0 & 0\\
 1 & 0 & 0
 \end{matrix}\right)\,,\\
  k^{a,1,b,1}=&k^{a,2,b,1}=9\left(\begin{matrix}
 0 & 0 & 0 \\
 0 & e^{-\pi i(2-\Gamma^2)/9} & 0\\ 
 0 & 0 & e^{-4\pi i(2-\Gamma^2)/9}
 \end{matrix}\right)\\
  k^{a,1,b,2}=&k^{a,2,b,1}=9\left(\begin{matrix}
 0 & 0 & 0 \\
 0 & 0 & e^{-2\pi i(2-\Gamma^2)/9} \\ 
 0 & e^{-5\pi i(2-\Gamma^2)/9} & 0
 \end{matrix}\right)\,,\\
		 k^{a,0,b,0}=&9\left(\begin{matrix}
 	0 & 1 & 1 \\
 	1 & e^{-\pi i(2-\Gamma^2)/9} & e^{-2\pi i(2-\Gamma^2)/9} \\ 
 	1 & e^{-5\pi i(2-\Gamma^2)/9} & e^{-4\pi i(2-\Gamma^2)/9}
 \end{matrix}\right)\,,
	 \end{split}
 \end{align}
These matrices can be obtained as follows:
When $r\cdot s=0$, then $k^{a,r,b,s}$ is identical to the result that we obtained for standard embedding.
Furthermore, modular invariance requires that the values in $k^{a,0,b,0}$ satisfy
\begin{align}
	k^{a,0,b,0}=e^{\pi i\frac{a^2}{9}(2-\Gamma^2)}k^{a,0,a+b,0}\,,\quad k^{a,0,b,0}=e^{-2\pi i\frac{ab}{9}(2-\Gamma^2)}k^{b,0,-a,0}\,,
\end{align}
where $a,b$ are considered modulo three~\cite{Henningson:1996jz}.
The relation
\begin{align}
	k^{a,0,b,0}=\sum\limits_{\stackrel{r,s}{rs>0}}k^{a,r,b,s}\,,
\end{align}
together with the fact that the matrices differ from those for the standard embedding only by a phase, then fixes all of the other coefficients.
The lattice sum involving the winding and momenta modes are given in each case by,
\begin{eqnarray} \label{gamma22}
\Gamma_{2,2}^{(r,s)} (\tau, \bar \tau) 
&=& \sum_{\stackrel{m_1, m_2, n_2 \in \mathbb{Z} }{n_1= \mathbb{Z} + \frac{r}{3} } }
q^\frac{p_L^2}{2} \bar q ^{\frac{p_R^2}{2}} e^{2\pi i m_1 s /3}\,, 
\end{eqnarray}
where the left- and right-moving momenta are given by
\begin{align}
	\begin{split}
\frac{1}{2} p_R^2 =& 
\frac{1}{2T_2 U_2} |-m_1 U + m_2 + n_1 T + n_2 TU |^2 \,, \\
\frac{1}{2}p_L^2 =& \frac{1}{2} p_R^2 + m_1n_1 + m_2 n_2\,,
 \label{plpr}
	\end{split}
\end{align}
with $T, U$ being K\"{a}hler and complex structure of $T^2$.

One can then define the different components of the new supersymmetric index ${\cal Z}_{\rm new}^{r,s}$ via
\be
{\cal Z}_{\rm new}(\tau,\bar \tau)=-\frac{1}{\eta^{24}(\tau)}\sum_{r,s=0}^2{\cal Z}_{\rm new}^{r,s}(\tau)\Gamma_{2,2}^{(r,s)}(\tau,\bar \tau).
\ee
We note that ${\cal Z}_{\rm new}^{r,s}$ are good modular functions and for $N=3$ orbifolds satisfy
\begin{align}
{\cal Z}_{\rm new}^{0,1}={\cal Z}_{\rm new}^{0,2}\,, \quad {\cal Z}_{\rm new}^{1,1}={\cal Z}_{\rm new}^{2,2}\,,\quad {\cal Z}_{\rm new}^{1,2}={\cal Z}_{\rm new}^{2,1}\,.
\end{align}
The value of $N_h-N_v$ can be extracted from the new supersymmetric index as
\be
N_h-N_v=-\frac{1}{4}\frac{1}{\eta^{20}(\tau)}\sum_{0,s=0}^2{\cal Z}_{\rm new}^{0,s}\big|_{q^{1/6}}\,,
\ee
where ${\cal Z}_{\rm new}^{0,0}=\frac{4}{3}E_4E_6$.
The results for the ${\cal Z}_{\rm new}^{r,s}$ sectors can be given in terms of modular forms for $\Gamma_1(3)$ and read
\begin{eqnarray}
{\cal Z}_{\rm new}^{0,1}&=&\frac{4}{3}\left(\hat a E_4E_6+\hat b {\cal E}_3^2(\tau)E_6+\hat c E_4(E_6+3{\cal E}_3(\tau)E_4)+\hat d {\cal E}_3^2(\tau)(E_6+3{\cal E}_3(\tau)E_4)\right),\nn\\
{\cal Z}_{\rm new}^{r,rk}&=&\frac{4}{3}\left(\hat a E_4E_6+\frac{\hat b}{9} {{\cal E}'_3}^2(\tau)E_6+\hat c E_4(E_6-{{\cal E}_3'}E_4)+\frac{\hat d}{9} {{\cal E}_3'}^2(E_6-{{\cal E}_3'}E_4)\right),
	\label{eqn:nssi}
\end{eqnarray}
where we have introduced ${\cal E}_3'(\tau)={\cal E}_3(\frac{\tau+k}{3})$ and the different values of $\hat a,\hat b,\hat c,\hat d$ are given in table~\ref{znew3}.
We note that the difference between the number of hypers and vectors in the standard embedding is as expected from earlier results in~\cite{Chattopadhyaya:2016xpa}.
\begin{table}[H]
	\renewcommand{\arraystretch}{0.5}
	\begin{center}
		\vspace{0.5cm}
		\begin{tabular}{|c|c|cccc|}
			\hline
			& & & & & \\
			Shift & $N_h-N_v$  & $\hat a$ & $\hat b$ & $\hat c$ & $\hat d$\\ \hline
			& & & & & \\
			$\frac13(1,-1,0^6;0^8)$& $-134$ & 0&0&$\frac{1}{4}$&0\\
			& & & & & \\
			$\frac13(2,1^4,0^3;0^8)$& $-80$ &$\frac{1}{16}$&$-\frac{9}{16}$&$-\frac{3}{16}$&$\frac{9}{16}$\\
			& & & & & \\
			$\frac13(2,0^7;2,0^7)$&  64 &$\frac{-1}{48}$  &$\frac{3}{16}$ &$\frac{1}{48}$ &$\frac{3}{16}$ \\
			& & & & & \\
			$\frac13(1,-1,0^6;2,1^2,0^5$)& 28 &0 &0 &$\frac{-1}{32}$ &$\frac{9}{32}$\\
			& & & & & \\
			$\frac13(2,1^2,0^5;2,1^4,0^3$)& 82 &$\frac{-1}{32}$ &$\frac{9}{32}$ &$\frac{3}{64}$ &$\frac{9}{64}$\\
			& & & & & \\
			\hline
		\end{tabular}
	\end{center}
	\vspace{-0.5cm}
	\caption{${\cal Z}^{r,s}_{\rm new}$ for different embeddings with $K3$ as $T^4/\mathbb{Z}_3$ and $N=3$ CHL orbifold.}\label{znew3}
	\renewcommand{\arraystretch}{0.5}
\end{table}
As we review in section~\ref{ssec:abriefreview}, the difference $N_h-N_v$ directly encodes the Euler characteristic $\chi$ of a dual Calabi-Yau compactification space via
\begin{align}
	\chi=2(N_v-N_h)+8\,.
\end{align}
Let us note again that when counting the vector multiplets $N_v$ we only consider those that originate from the $E_8\times E_8$ and not the generic four vectors that arise from the $T^2$.
It turns out that the coefficients of the new supersymmetric indices for all of the four non-standard embeddings can be expressed as
\begin{align}
	\begin{split}
	\hat{a}=&\frac{1}{2^73^3}(48+\chi)\,,\quad\hat{b}=-\frac{1}{2^73}{48+\chi}\,,\\
	\hat{c}=&-\frac{1}{2^83^3}(456+5\chi)\,,\quad\hat{d}=\frac{1}{2^83}(264+\chi)\,.
	\end{split}
	\label{eqn:abcdchi}
\end{align}
This would imply the values
\begin{align}
	(\hat{a},\,\hat{b},\,\hat{c},\,\hat{d})=\left(\frac{3}{32},\,-\frac{27}{32},\,-\frac{17}{64},\,\frac{45}{64}\right)\,,
	\label{eqn:abcdstandard}
\end{align}
for the standard embedding with $N_h-N_v=-134$, which is clearly different from what we found via an explicit calculation.
Note that for compactifications on $K3\times T^2$ and $(K3\times T^2)/\mathbb{Z}_2$ it was found that the new supersymmetric index can
be expressed in terms of the Euler characteristic for \textit{all} of the standard and non-standard embeddings that had been considered~\cite{Stieberger:1998yi,Chattopadhyaya:2016xpa}.
It would be interesting to understand what makes the standard embedding on $(K3\times T^2)/\mathbb{Z}_3$ different and whether a non-standard embedding exists that realizes~\eqref{eqn:abcdstandard}.

\section{Heterotic computation of Gopakumar-Vafa invariants}\label{hetgv}
In this section we compute the Gopakumar-Vafa invariants from the heterotic side for the non-standard embeddings discussed above.
The Gopakumar-Vafa invariants $n^g_\beta$ correspond to weighted sums of multiplicities of BPS states that arise from M2-branes wrapping curves of class $\beta\in H_2(M)$ in a Calabi-Yau threefold $M$~\cite{Gopakumar:1998ii,Gopakumar:1998jq}.
We will loosely refer to $n^g_\beta$ as Gopakumar-Vafa invariants of genus $g$ because they only receive contributions from M2-branes that wrap curves with genus less than or equal to $g$.
To obtain the invariants one needs to calculate the gravitational coupling in the low energy ${\cal N}=2$ effective theory.
The main ingredients of the calculation are different sectors of the new supersymmetric index.

At certain points in the moduli space one encounters singularities in the gravitational coupling.
Towards the end of this section we shall discuss these singularities known as conifold singularities.
The observations here are very similar to the ones obtained in~\cite{Chattopadhyaya:2017zul}.
The strength at these singular points are given by the Gopakumar-Vafa invariants at these points.
It is also similar to the results for the standard embedding in~\cite{Chattopadhyaya:2017zul} that the untwisted sector is devoid of these singular points.

\subsection{Extraction of Gopakumar-Vafa invariants from $F_g$}
The gravitational couplings $F_g$ of the low energy effective ${\cal N}=2$ theories appear as
\begin{equation}
S = \int F_g(y,  \bar y) \cdot F_+^{2g-2} R_+^2\,,
\end{equation}
where $F_+,\, R_+$ are the self-dual parts of the graviphoton and Riemann tensor and $y,\bar y$ denote the dependence on vector moduli.

For $K3\times T^2$ without any orbifolding, the couplings have been computed from the corresponding heterotic one-loop amplitude in~\cite{Marino:1998pg} using the so-called Borcherd lift~\cite{Borcherds:1996uda}.
The calculation without any graviphoton insertion, which leads to the prepotential of the supergravity theory, was done earlier in~\cite{Harvey-Moore}.
For compactifications on $(K3\times T^2)/\mathbb{Z}_3$ with standard embeddings the calculation has been performed in~\cite{Chattopadhyaya:2017zul}.

Through the computation of this $F_g$ we can obtain predictions for Gopakumar-Vafa invariants at genus $g$ of a dual Calabi-Yau threefold.
The essential ingredient in computing the $F_g$ is given by the new supersymmetric index. Under standard embedding for any $g'$ orbifold of $K3$ the results of this index were given in \cite{Chattopadhyaya:2016xpa}
\begin{equation}
\mathcal Z_{new}(q,\bar q)=-4 \sum_{r,s=0}^{N-1}\Gamma_{2,2}^{(r,s)} f^{(r,s)}(\tau)\,,
\end{equation}
where one introduces
\begin{align}\label{defcrsfrs}
f^{(r,s)} (\tau) =&  \frac{1}{2\eta^{24} (\tau) } 
E_4 
\left[ \frac{1}{4} \alpha_{g'}^{(r,s)} E_6  - 
\beta_{g'}^{(r,s)} (\tau) E_4 \right] =\sum_{ l \in \frac{\mathbb{Z}}{N} }  c^{(r, s)} ( l) q^{l}\,.
\end{align}
The $\alpha_{g'}^{(r,s)}$ are numerical constants while $\beta_{g'}^{(r,s)}$ are weight 2 modular forms for $\Gamma_1(N)$, where $N$ corresponds to the order of the orbifold group.\\

It has previously been shown~\cite{Antoniadis:1992sa,Antoniadis:1995zn,Antoniadis:1992rq,Antoniadis:1993ze} that the one-loop integral $F_g$ is
\begin{eqnarray} \label{fgexp}
F_g &=& \frac{1}{ 2\pi^2( g!) ^2} \int \frac{ d^2 \tau}{\tau_2} \left\{  \frac{1}{\tau_2^2 \eta^2(\tau)}
{\rm Tr}_R
\left[   ( i \bar \partial X)^ {( 2g - 2) } 
(-1)^{F} F q^{L_0 - \frac{c}{24} } \bar q^{\tilde{L}_0 - \frac{\tilde{c}}{24} } \right ]  \right. \\ \nonumber
&& \qquad \qquad \qquad  \left.  \langle \prod_{i=1}^g \int d^2 x_i Z^1 \partial Z^2 (x_i)  
\prod_{j = 1}^g \int d^2 y_j \bar Z^2  \partial \bar Z^1 ( y_j)  \rangle  \right\}.
\end{eqnarray}
Here $X$ is the complex coordinate on the $T^2$ and $Z^1, Z^2$ are the complex
coordinates of the transverse non-compact space time.
The above result is obtained from the Wick contractions of the graviton and graviphoton vertex operators.
Factorizing the non-compact directions, the internal CFT trace is given by
\begin{eqnarray} \label{intrace}
&& \frac{1}{\eta^2(\tau) } {\rm Tr}_R
\left[   ( i \partial X)^{( 2g - 2)}  
(-1)^{F} F q^{L_0 - \frac{c}{24} } \bar q^{\tilde{L}_0 - \frac{\tilde{c}}{24} } \right ]
= \\ \nonumber 
& & \qquad\qquad\qquad\quad 2 \frac{1}{\eta^{24} (\tau)}  
\sum_{r,s} \Gamma_{2,2}^{(r,s)} \left(\frac{(p_R^{(r, s)})} {\sqrt{2T_2 U_2} } 
\right)^{(2g -2) } q^{|p_L|^2/2 }\bar q^{|p_R|^2/2} f^{r,s}(\tau)\,.
\end{eqnarray}
The only essential difference between this trace and the new supersymmetric index computation are the extra powers of $p_R^{(r,s)}$.
It is evident that under non-standard embedding $f^{(r,s)}$ would have to be replaced by $\frac 1 {4 \eta^{24}(\tau)} {\cal Z}_{\rm new}^{r,s}$.
The momenta $p_L^{(r,s)}, p_R^{(r,s)}$ are given in equation~\eqref{plpr}, assuming no other moduli dependence on the heterotic side.

In order to make contact with the enumerative geometry of a type II Calabi-Yau compactification that produces the same effective action
we need to extract the holomorphic part of the integral $F_g$.
The purely holomorphic contribution to the gravitational couplings are given by
\begin{align}
	\begin{split}
	\bar{F}_g^{{\rm hol}}=&\frac{(-1)^{g-1}}{\pi^2} \sum_{s=0}^{N-1} \left( \sum_{m>0}e^{-2\pi i n_2s/N}c^{(r,s)}_{g-1}(n_1n_2/N){\rm Li}_{3-2g}(e^{2\pi im\cdot y})\right.\\
	&\left.+\frac{1}{2}c_{g-1}^{(0,s)}(0)\zeta(3-2g) \right)\,.
	\end{split}
\label{fhol}
\end{align}
where $y=(T,U)$ with $T,U$ being the complex and the complexified K\"ahler structure of $T^2$. 
The coefficients $c^{(r,s)}_{g-1}(n_1n_2/N)$ are obtained from
\begin{equation}\label{defcg}
\frac{1}{4\eta^{24}}{\cal Z}_{\rm new}^{r,s}{\cal P}_{2k +2} (G_2,G_4,\dots,G_{2g}) 
=\sum_{l \in \frac{\mathbb{Z}}{N} } c_{g-1}^{(r,s)} (l) q^l \,,
\end{equation}
where $G_{2k}=2 \zeta(2k) E_{2k}$, $\zeta$ is the Riemann zeta function and $E_{2k}$ denote the Eisenstein series of weight $2k$.
${\cal P}_{2g}$ is defined through  the Schur polynomial of order $g $ in the following manner
\begin{equation}
{\cal P}_{2g} ( x_1, x_2, \cdots x_g) = - {\cal S} \left( x_1, \frac{1}{2} x_2, \cdots \frac{1}{g} x_g\right)\,.
\end{equation}
For an orbifold of order $N$ the lattice sum $m>0$ involves the points
\begin{eqnarray}
	m=(n_1, n_2) \geq 0\,, \quad\hbox{but } ( n_1, n_2 ) \neq ( 0, 0 )\,, \\ \nonumber
( r/N, - n_2) \,, \qquad \hbox{with}\;\;  n_2>0 \;\;\hbox{and}\;\; rn_2 \leq N\,.
	\label{eqn:mg0}
\end{eqnarray}
The Gopakumar-Vafa invariants of the potential Calabi-Yau duals can now be extracted from the heterotic results by first writing down the Gopakumar-Vafa form of the genus $g$ topological amplitude.
In terms of Gopakumar-Vafa invariants $n_m^g$ and for $g>1$ it can be written as~\cite{Gopakumar:1998ii,Gopakumar:1998jq}
\bea \label{fgv}
F^{{\rm GV}}_g&=&\frac{(-1)^g |B_{2g}B_{2g-2}|\chi(X)}{4g (2g-2)(2g-2)!}\\ \nn
&+& \sum_{\beta}\left[\frac{|B_{2g}|n_m^0}{2g (2g-2)!}+\frac{2(-1)^g n_m^2}{(2g-2)!}\pm...-\frac{g-2}{12}n_{m}^{g-1}+n_m^g\right]{\rm Li}_{3-2g}(e^{2\pi i m\cdot y})\,,
\eea
where $B_{2g}$ are the Bernoulli numbers and  $\chi(X)$ is the Euler characteristic of the Calabi-Yau.
For $g=0$ we get 
\begin{equation} \label{fg0gv}
F_0^{\rm GV} = \zeta(3) \frac{\chi(X)}{2} + \sum_{m>0} n_{m}^0 {\rm Li}_3( e^{2\pi i m\cdot y})\,.
\end{equation}
For $g=1$ we have
\begin{equation} \label{f1gv}
F_1^{\rm GV} = \sum_{m>0} \left( \frac{1}{12} n_m^0 + n_m^1 \right) {\rm Li}_{1}  
( e^{2\pi i m\cdot y})\,.
\end{equation}
Comparing the constant term of (\ref{fhol}) and (\ref{fgv}) gives the following relation between the toplogical amplitudes
\begin{equation}\label{relfgfhol}
F_g^{{\rm GV}} = \frac{(-1)^{g+1}}{2(2\pi )^{2g-2}} \bar F_g^{\rm hol}\,.
\end{equation}
By proceeding recursively, one obtains expressions for the Gopakumar-Vafa invariants of lowest genus
\begin{align}
\begin{split}
n^{0}_{(n_1,n_2)}=&2 \sum_{s=0} ^{N-1} e^{-\frac {2\pi n_2s} N} c_{-1}^{(r,s)} \,,\\
n^1_{(n_1,n_2)} =& \frac 1 {2 \pi ^2} \sum _{s=0} ^N e^{-\frac {2\pi i n_2 s} N} c_0^{(r,s)}(m^2/2) - \frac 1 {12} n^{0}_{(n_1,n_2)}\,, \\
n^2 _{(n_1,n_2)} =& \frac 1 {8 \pi^4} \sum_{s=0}^{N-1} e^{-\frac{2\pi i n_2 s}N} c_1^{(r,s)}(m^2/2)-\frac{|B_4|} 8 n^0_{n_1,n_2}\,,	
\end{split}
\end{align}
where $r=n_1N$ mod $N$ and $m^2=2n_1n_2$. 
\subsection{Conifold singularities}
Conifold singularities are reflected in the gravitational thresholds as poles where vectors become massless.
The polylogarithm ${\rm Li}_a(z)$ is singular at $z=1$ and the leading behaviour around the pole is for $a<0$ given by
\be
{\rm Li}_a(z)=(z\frac{d}{dz})^{|a|}\left. \frac{1}{1-z}\right|_{z\rightarrow 1} \sim
a!\frac{1}{(1-z)^{|a|+1}}\,.
\ee
Hence the leading divergence for $g>1$ in $\bar F_g^{\rm hol}$ can be written as
\be
\bar F_g^{\rm hol}|_{m\cdot y \rightarrow 0 } \sim 
\frac{(-1)^{g-1}}{\pi^2 }
\sum_{s=0}^{ N-1}
e^{-2\pi i n_2s/N}c^{ (r,s)}_{g-1}(n_1n_2)(2g-3)!\frac{1}{\{1-e^{2\pi i m\cdot y}\}^{2g-2}}\,,
\ee
where
\be
m\cdot y=n_1T_1+n_2U_1=0\,,\qquad n_1T_2+n_2U_2=0\,.
\ee
Coefficients of those functions are related to the Euler characteristics however and can be computed from the spectrum or ${\cal Z}_{\rm new}$.
Since $T_2, U_2>0$, being the imaginary parts of the K\"ahler and complex structure of the torus $T^2$, and $m$ is constrained by~\eqref{eqn:mg0} we see that $n_2$ is negative at these singularities. 
Therefore in the untwisted sector a possible  singularity  lies at
\begin{equation}
(n_1,n_2)  = ( 1, -1) \,, \qquad n_1n_2 = -1\,,
\end{equation}
However, a close inspection of the new supersymmetric index shows that for any standard or non-standard embedding for $N=3$ CHL orbifolds there lies no conifold singularity in the untwisted sector,
This was previously observed for the standard embedding results in \cite{Chattopadhyaya:2017zul}. 
 
\section{Dual type IIA compactifications on Calabi-Yau threefolds}
\label{sec:iiaduals}
In the previous sections we have discussed in detail the calculation of the gravitational coupling
\begin{align}
	S=\int F_g(y,\bar{y})F_+^{2g-2}R_+^2\,,
	\label{eqn:FRcoupling}
\end{align}
for heterotic $E_8\times E_8$ compactifications on $(K3\times T^2)/\mathbb{Z}_3$.
Following the arguments that we review below, one expects that there exists a dual type IIA compactification on a genus one fibered Calabi-Yau threefold with three-sections that also exhibits a $K3$ fibration.

The calculations of the heterotic one-loop calculation in this paper extends earlier work~\cite{Chattopadhyaya:2016xpa,Chattopadhyaya:2017zul} in that we consider several non-standard embeddings of the spin connection into the gauge connection.
In particular, two of the corresponding effective theories exhibit a spectrum such that the gauge group can be maximally higgsed to $U(1)^3$.
We can therefore hope that dual type IIA compactifications on Calabi-Yau threefolds to these models exist such that the geometries are genus one and $K3$-fibered but have the lowest possible number of K\"ahler moduli $h^{1,1}=3$.

In this section we will systematically engineer such Calabi-Yau manifolds using tools from toric geometry and show that
they are indeed dual to heterotic compactification on $(K3\times T^2)/\mathbb{Z}_3$.
To this end we employ the modular bootstrap and obtain all-genus topological string amplitudes on the corresponding geometries.
We then match these with the results from the heterotic one-loop calculation.

\subsection{A brief review of heterotic/type II duality}
\label{ssec:abriefreview}
Let us briefly review the basics of heterotic/type II duality and why the Calabi-Yau manifolds require a particular fibration structure. 
A standard reference on this duality is still~\cite{Aspinwall:1996mn} and a more recent discussion can be found in~\cite{Braun:2016sks}.

In six dimensions, heterotic strings on $T^4$ are dual to type IIA strings on a $K3$~\cite{Hull:1994ys,Witten:1995ex}.
The four dimensional duality between heterotic strings on $K3\times T^2$ and type IIA strings on a Calabi-Yau threefold~\cite{Kachru:1995wm} can then be motivated by an adiabatic argument~\cite{Vafa:1995gm}.
More precisely, one can fiber the geometries on both sides of the duality over a $\mathbb{P}^1$ such that the fibration of $K3$ over $\mathbb{P}^1$ is a Calabi-Yau threefold.
Matching the amount of unbroken supersymmetries requires the fibration on the heterotic side to only involve a 2-torus inside $T^4$ such that the resulting geometry is $K3\times T^2$.
From the perspective of the type II compactification, the heterotic string arises from a 5-brane that wraps the generic $K3$ fiber.
The adiabatic argument can fail at points where the $K3$ fiber degenerates and this can be related to the presence of 5-branes on the heterotic side of the duality~\cite{Braun:2016sks}. 

The heterotic compactification on $K3\times T^2$ exhibits a self-duality group which contains the $SL(2,\mathbb{Z})$ action on the complex structure of the torus.
After taking a $\mathbb{Z}_N$ quotient which acts by an order $N$ shift on the torus, this duality group is broken down to the congruence subgroup $\Gamma_1(N)\subset SL(2,\mathbb{Z})$ that
preserves the generator of a $\mathbb{Z}_N$ subgroup of $T^2$~\cite{Vafa:1995gm,Cota:2019cjx}.
The duality group of the heterotic string is realized on the type IIA side as monodromies in the stringy K\"ahler moduli space of the Calabi-Yau threefold~\cite{Klemm:1995tj,Kachru:1995fv}.
This implies that for heterotic strings on $(K3\times T^2)/\mathbb{Z}_N$ the dual Calabi-Yau compactification space should be a genus one fibration with $N$-sections~\cite{Schimannek:2019ijf,Cota:2019cjx}. 

The moduli space of the four-dimensional ${\cal N}=2$ effective theories factorizes into
\begin{align}
	\mathcal{M}=\mathcal{M}_{vec.}\times\mathcal{M}_{hyp.}\,,
\end{align}
where $\mathcal{M}_{vec.}$ is parametrized by the expectation values of scalar fields from vector multiplets and $\mathcal{M}_{hyp.}$ is correspondingly parametrized by scalars from hyper multiplets.
The coefficients $F_g(y,\bar{y})$ in~\eqref{eqn:FRcoupling} only depend on the vector moduli.

On the heterotic side there are four massless vector fields from reducing the metric and the B-field along the $T^2$ of which one contributes to the single graviton multiplet.
The remaining three vector fields combine with 6 real scalars that arise from the dual of the B-field in four dimensions, the dilaton and the Kaluza-Klein modes of the metric and the B-field along $T^2$, as well as an appropriate number of fermions, into three vector multiplets.
The dilaton and the dual of the B-field combine into a complex axio-dilaton while the Kaluza-Klein modes of metric and B-field can be interpreted as the complex structure and the complexified K\"ahler structure of $T^2$.
Additional massless vector multiplets might descend from the 10-dimensional $E_8\times E_8$ gauge group but those can acquire a mass from the choice of gauge background.

On the type IIA side, the scalars in the vector multiplets arise as Kaluza-Klein modes of the K\"ahler form and the B-field, both of which can be expanded in harmonic 2-forms that in turn correspond to $H^{1,1}(M)$ classes of the Calabi-Yau $M$.
The graviphoton on the other hand corresponds to the Ramond-Ramond one-form.
This implies that for a heterotic compactification that leads to the minimum of four vector fields a dual Calabi-Yau compactification space needs to exhibit $h^{1,1}=3$.
The three minimal K\"ahler moduli of the Calabi-Yau can be related to the volume of the generic fiber, the volume of the base of the $K3$ fiber and the volume of the $\mathbb{P}^1$ base of the $K3$ fibration.
Moreover, the coefficients $F_g(y,\bar{y})$ are topological~\cite{Antoniadis:1993ze,Antoniadis:1995zn} and can be interpreted as genus $g$ amplitudes of the A-twisted topological string theory on $M$.

Under the duality, the complex structure and the complexified K\"ahler structure of the $T^2$ on the heterotic side can be identified with the complexified volumes of fiber and base of the generic $K3$ fiber.
The heterotic axio-dilaton is identified with the complexified volume of the $\mathbb{P}^1$ base of the $K3$ fibration.
On the heterotic side one can T-dualize along the cycle that is not involved in the $\mathbb{Z}_N$ quotient and this exchanges complex and complexified K\"ahler structure on $T^2$.
This duality is also realized as a monodromy in the complexified K\"ahler moduli space of the dual Calabi-Yau.

\subsection{Constructing Calabi-Yau duals}
We now want to construct Calabi-Yau threefolds that exhibit both a $K3$ fibration and a genus one fibration while $h^{1,1}=3$.
To this end let us first introduce some mathematical background about $K3$ surfaces~\cite{Braun:2014oya,huybrechts_2016}. 

In constrast to the wealth of topologically distinct Calabi-Yau threefolds there is only one topological type of $K3$ surfaces.
The non-vanishing Hodge numbers are always $h^{0,0}=h^{2,0}=h^{0,2}=h^{2,2}=1$ and $h^{1,1}=20$.
However, for a given $K3$ surface $S$, the $H^2(S,\mathbb{Z})$ cohomology decomposes into the \textit{Néron-Severi lattice}
\begin{align}
	\text{NS}_{K3}=H^2(S,\mathbb{Z})\cap H^{1,1}(S)\,,
\end{align}
and the orthogonal complement $T$ which is called the \textit{transcendental lattice}.
The rank of $\text{NS}_{K3}$ is called the \textit{Picard number} $\rho(S)$.
A $K3$ fibration is called $\Lambda_S$ \textit{polarized} if there is a set of divisors on the total space $M$ of the fibration such that the restrictions of the divisors to a generic fiber
generate a sublattice $\Lambda_S\subset\text{NS}_{K3}$.
For $K3$ surfaces of Picard number two one can write the intersection form on $\text{NS}_{K3}$ as
\begin{align}
	I=\left(\begin{array}{cc}2a&b\\b&2c\end{array}\right)\,,\quad a,b,c\in\mathbb{Z}\,,\quad 4ac-b^2<0\,.
\end{align}
Moreover, the surface is genus one fibered with an $N$-section if and only if $b^2-4ac=N^2$.

Our goal in this section is to construct genus one fibered Calabi-Yau threefolds that also exhibit a $K3$ fibration and have $h^{1,1}=3$.
Let us first note that every genus one fibration that exhibits $1$-, $2$- or $3$-sections is birational to a hypersurface in a fibration of weighted projective spaces~\cite{Braun:2014oya}.
The respective fibers are $\mathbb{P}(1,2,3)$, $\mathbb{P}(1,1,2)$ and $\mathbb{P}^2$.
Moreover, the base of a genus one fibration can only be $\mathbb{P}^2$, a Hirzebruch surface $\mathbb{F}_m$, the Enriques surface or blowups of these geometries~\cite{Grassi1991}.
A fibration over $\mathbb{P}^2$ will never exhibit a $K3$ fibration and, since we are interested in Calabi-Yau manifolds with $h^{1,1}=3$, the only bases
that we have to consider are Hirzebruch surfaces $\mathbb{F}_m,\,m=0,1,2$.
If we are only interested in genus one fibrations with $N$-sections for $N\le 3$ we can therefore restrict ourselves to Calabi-Yau threefolds that are hypersurfaces in toric ambient spaces.
Moreover, since the base of the $K3$ fibration has to be $\mathbb{P}^1$, the fibration has to arise from a compatible toric fibration of the ambient space.
The generic $K3$ fiber will therefore also be a hypersurface in a three-dimensional toric variety.

Six $K3$ surfaces with Picard number two that exhibit a genus one fibration with $N$-sections are realized as generic hypersurfaces in three-dimensional toric ambient spaces.
Two more cases have non-toric sections that are combined into a non-toric ``pseudo'' $N$-section.
Those fibers can still lead to fibrations with $h^{1,1}=3$ when the components of the pseudo $N$-section experience monodromy along the $\mathbb{P}^1$ base of the threefold.

Given a three dimensional reflexive polytope that corresponds to a $K3$ we can construct a four-dimensional polytope by lifting $\Delta^{K3}$ into $\mathbb{Z}^3\times \mathbb{Z}$ and adding the points
\begin{align}
	\nu'_1=(\nu,1)\,,\quad\nu'_2=(0,-1)\,.
	\label{eqn:lift}
\end{align}
The convex hull is reflexive whenever $\nu\in 2\Delta^{K3}$~\cite{Braun:2014oya}.
If the four dimensional polytope admits a fine regular star triangulation that is compatible with the fibration then the generic Calabi-Yau hypersurface in the corresponding toric ambient space
will be a $K3$ fibration.

\paragraph{Elliptic fibrations}
There is only one case with a $1$-section which is the degree twelve hypersurface in $\mathbb{P}(1,1,4,6)$.
The intersection form on the Néron-Severi lattice is
\begin{align}
	U=\left(\begin{array}{cc}0&1\\1&0\end{array}\right)\,.
\end{align}
Let us write the points of the polytope $\Delta^{(1)}$ as
\begin{align}
\begin{blockarray}{crrr}
\begin{block}{c(rrr)}
	\nu^1& 1& 0& 0\\
	\nu^2& 0& 1& 0\\
	\nu^3&-2&-3& 0\\
	\nu^4&-2&-3& 1\\
	\nu^5&-2&-3&-1\\
\end{block}
\end{blockarray}\,,
\label{eqn:k3points1}
\end{align}
There are three more points in the interior of facets and they correspond to divisors that do not intersect the generic hypersurface.
In general those points will not lift to points in the interior of facets of the four-dimensional polytope.
The only inequivalent choices of points $\nu\in 2\Delta^{(1)}$ that lead to a Calabi-Yau threefold with $h^{1,1}=3$ are
\begin{align}
	\nu=(-4,-6,-n, 1)\,,
\end{align}
for $n\in\{0,1,2\}$.
They lead to the well-known elliptic fibrations over $\mathbb{F}_n$.
The Gopakumar-Vafa invariants of degree zero with respect to the $\mathbb{P}^1$ base of the $K3$ fibration are the same for all three geometries.
Nevertheless, they differ already in the triple intersection numbers and correspond to heterotic compactifications on $K3\times T^2$ with instantons distributed as $(12+n,12-n)$ among the two $E_8$'s.

\paragraph{Genus one fibrations with three-sections}
Three $K3$ surfaces that exhibit a genus one fibration with three sections are realized as generic hypersurfaces in three-dimensional toric ambient spaces.
They correspond to the reflexive polytopes that are the convex hull $\Delta_n^{(3)}$ of the points
\begin{align}
\begin{blockarray}{crrr}
\begin{block}{c(rrr)}
	\nu^1& 1& 0& 0\\
	\nu^2& 0& 1& 0\\
	\nu^3&-1&-1& 0\\
	\nu^4& n& 0& 1\\
	\nu^5& 0& 0&-1\\
\end{block}
\end{blockarray}\,,
\label{eqn:k3points}
\end{align}
with $n\in\{-1,0,1\}$.
The intersection forms on the Néron-Severi lattice are
\begin{align}
	I_{n}^{(3)}=\left(\begin{array}{cc}
		2(1+n)&3\\3&0
	\end{array}\right)\,.
	\label{eqn:insnst}
\end{align}
For $n=2$ the corresponding $K3$ surface has Picard rank four but only two generators of $\text{NS}_{K3}$ descend from toric divisors on the ambient space.
One of the toric divisors corresponds to a ``pseudo'' three-section with trivial monodromy acting on the intersections with the generic fiber.
However, if this geometry is fibered over a $\mathbb{P}^1$, the pseudo three-section can turn into a real three-section.
The fibration is then only polarized with respect to a sublattice of the generic fiber.
The intersection form $I^{(3)}_2$ with respect to the toric divisors on the $K3$ is equivalent to $I^{(3)}_{-1}$.

Again we can lift $\Delta_n^{(3)}$ into $\mathbb{Z}^3\times\mathbb{Z}$ and add the points~\eqref{eqn:lift},
\begin{align}
	\nu'_1=(\nu,1)\,,\quad\nu'_2=(0,-1)\,.
\end{align}
The convex hull is reflexive whenever $\nu\in2\Delta^{(3)}_n$.
Furthermore, we can use lattice automorphisms to impose $\nu_3\le 0$.
This allows us to identify the divisor that corresponds to the point $\nu'_2$ as the vertical divisor $\pi^{-1}(B)$ and the divisor corresponding to $\nu^5$ as $\pi^{-1}(F)$, where $B$ and $F$ are the base and the fiber of the Hirzebruch surface that is the base of the genus one fibration.
All of the corresponding Calabi-Yau threefolds will have $h^{1,1}=3$ except if $n=2$ and $\nu$ takes one of the three values
\begin{align}
	\nu=(0,0,-2)\,,\quad\nu=(1,0,-1)\,,\quad=(2,0,0)\,.
\end{align}
For all choices of $n$ and $\nu$ we can express the Euler characteristic $\chi_n^{(3)}(\nu)$ of the Calabi-Yau threefold in terms of the vector $\nu$ in~\eqref{eqn:lift},
\begin{align}
	\chi_n^{(3)}(\nu)=n(-12\nu_1+6\nu_2)+6n^2\nu_3-144\,.
\end{align}

It turns out that it is exactly the fibrations constructed from the fibers with $n=-1$ and $n=2$, i.e. those where the intersection form on the polarization lattice is
\begin{align}
	I=\left(\begin{array}{cc}
		0&3\\3&0
	\end{array}\right)\,,
	\label{eqn:intst}
\end{align}
that can be matched with weakly coupled heterotic compactifications on $(K3\times T^2)/\mathbb{Z}_3$.
When $n\in\{0,1\}$, the fiber-base duality of the $K3$ fiber and therefore the T-duality of any heterotic dual is broken.
For hypersurfaces in toric ambient spaces, the  Gopakumar-Vafa invariants at genus zero can be calculated using standard techniques~\cite{Hosono:1993qy}. 
We find that for $n\in\{-1,2\}$ the invariants at genus zero and of degree zero with respect to the base of the $K3$ fibration only depend on the Euler characteristic and for low degrees we list the result in table~\ref{eqn:gv1}.
\begin{table}[h!]
{\tiny
\begin{align*}
\begin{array}{|c|ccccc|}
\hline d_B\backslash d_F&0&1&2&3&4\\\hline
	0&0 & \frac{\chi }{2}+240 & \frac{\chi }{2}+240 & -\chi  & \frac{\chi }{2}+240 \\
	1& \frac{\chi }{2}+240 & 1962-3 \chi  & \frac{10 \chi }{3}+18016 & \frac{15 \chi }{2}+95454 & 413280-30 \chi  \\
	2& 0&\frac{\chi }{2}+240 & \frac{10 \chi }{3}+18016 & 413280-30 \chi  & 54 \chi +5694624 \\
	3& 0&0&-\chi  & \frac{15 \chi }{2}+95454 & 54 \chi +5694624 \\
	4& 0&0&0&\frac{\chi }{2}+240 & 413280-30 \chi  \\
	5& 0&0&0&0&\\\hline
\end{array}
\end{align*}
}
	\caption{Genus zero Gopakumar-Vafa invariants of degree zero with respect to the base of the $K3$ fibration for geometries where the intersection form on the polarization lattice is~\eqref{eqn:intst}.
	This matches the heterotic prediction from~\eqref{eqn:nssi} with~\eqref{eqn:abcdchi}.}
	\label{eqn:gv1}
\end{table}
Here the degrees $d_B,d_F$ respectively correspond to the $\mathbb{P}^1$ base and the genus one fiber of the generic $K3$ fiber.
More precisely, for a given class $\beta\in H_2(M)$ one has
\begin{align}
	d_F=E_0\cdot \beta\,,\quad d_B=\pi^{-1}(B)\cdot \beta\,,
\end{align}
where $E_0$ is the divisor associated to the three-section.
This implies that $(d_B,d_F)=(3,0)$ is the class of the generic genus one fiber while degrees $d_F\equiv1,2\,\,\text{mod}\,\,3$ correspond to contributions from isolated $I_2$ singular fibers in special $K3$ fibers.
We find a total of $17$ inequivalent toric ambient spaces that lead to families of $K3$ fibered Calabi-Yau hypersurfaces with eight different Euler characterstics
\begin{align}
	\chi\in\{-192,\,-168,\,-156,\,-150,\,-144,\,-138,\,-132,\,-120\}\,.
\end{align}
Three models lead to $\chi=-192$ and are special in that the three-section is actually a union of three sections.
The number of K\"ahler moduli is then $h^{1,1}=5$.

It turns out that the invariants exactly match those predicted from~\eqref{fhol} with~\eqref{eqn:nssi} and~\eqref{eqn:abcdchi} when we identify $d_F=n_1,\,d_B=n_1+n_2$.
In other words, the complexified volume of the generic genus one fiber $\tau$ and the volume of the base of the $K3$ fiber $t$ can be related to the complex structure 
and the complexified K\"ahler structure $T,U$ on the heterotic side via
\begin{align}
	\tau=U\,,\quad t=T-U\,.
\end{align}
This identification is the same as for the original $STU$-model.
In summary, we find that a $K3$ fibered Calabi-Yau threefold exhibits Gopakumar-Vafa invariants with respect to the $K3$ fiber
that match the predictions from the corresponding one-loop amplitudes of  heterotic strings on $(K3\times T^2)/\mathbb{Z}_3$ if and only if the intersection form on
the polarization lattice is~\eqref{eqn:intst}.

Two of the heterotic compactifications with non-standard embedding on $(K3\times T^2)/\mathbb{Z}_3$ led to a maximally higgsable gauge group
and the corresponding predictions for the Euler characteristics were $\chi=-120$ and $\chi=-156$.
The geometries with $(n;\nu)$ given by $(-1;2,0,0)$ and $(2;-2,-2,0)$ have Euler characteristic $\chi=-120$.
However, both are genus one fibrations over Hirzebruch $\mathbb{F}_0$ and they turn out to be Wall equivalent.
All of their genus zero invariants appear to match and we therefore have a unique candidate Calabi-Yau dual.

On the other hand, the geometries with the data $(-1;0,2,0)$, $(-1;-1,0,0)$, $(-1;0,1,-1)$ and $(2;0,1,2)$ have Euler characteristic $\chi=-156$.
The former two are again Wall equivalent and fibered over $\mathbb{F}_0$ while the latter two are also Wall equivalent but fibered over $\mathbb{F}_1$.
Their intersection rings as well as the Gopakumar-Vafa invariants for non-zero degrees with respect to the base of the $K3$ fibration are different.
We therefore find a situation that is similar to the $STU$ models with instanton numbers $(12-n,12+n)$ for $n=0,1,2$.
They also produce the same new supersymmetric index but are respectively dual to generic elliptic fibrations over $\mathbb{F}_n,\,n=0,1,2$~\cite{Morrison:1996pp}.

Without the CHL orbifold, the instanton number of the non-standard embedding that leads to the predicted value $\chi=-156$ is $n=3$~\cite{Stieberger:1998yi}.
One might be tempted to divide this number by three and conjecture that the geometry with data $(-1;0,1,-1)/(2;0,1,2)$ is the correct dual.
However, determining the instanton number of the dual heterotic compactification directly from the Calabi-Yau geometry is difficult although a proposal was made in~\cite{Braun:2016sks}.
We leave a further investigation of this issue for future work.
Note that the instanton number without CHL quotient of the non-standard embedding that predicts $\chi=-120$ is $n=0$.
This seems compatible with the fact that the unique Calabi-Yau hypersurface in our classification that reproduces this number is fibered over $\mathbb{F}_0$.

The Euler characteristic of all threefolds from $n=0$ is $\chi_0^{(3)}(\nu)=-144$ and the Gopakumar-Vafa invariants at genus $0$ with respect to the $K3$ fiber given in table~\ref{eqn:gv2}
while for $n=1$ the invariants are listed in table~\ref{eqn:gv3}.
\begin{table}
{\tiny
\begin{align*}
\begin{array}{|c|cccccc|}
	\hline d_B\backslash d_F&0&1&2&3&4&5\\\hline
	0& 0 & \frac{\chi }{2}+240 & \frac{\chi }{2}+240 & -\chi  & \frac{\chi }{2}+240 & \frac{\chi }{2}+240 \\
	1& \frac{\chi }{2}+84 & 318-\chi  & 4776-\frac{\chi }{2} & 4 \chi +32262 & 156540-\frac{11 \chi }{2} & 665484-2 \chi  \\
	2& 0 & \frac{\chi }{2}+84 & 4776-\frac{\chi }{2} & 156540-\frac{11 \chi }{2} & \frac{37 \chi }{2}+2492400 & 26767452-\frac{17 \chi }{2} \\
	3& 0 & -2 & \frac{\chi }{2}+240 & 3 \chi +94806 & 5681624-27 \chi  & 48 \chi +159180264 \\
	4& 0 & 0 & 0 & 4776-\frac{\chi }{2} & \frac{37 \chi }{2}+2492400 & 223507008-84 \chi  \\
	5& 0 & 0 & 0 & 0 & 156540-\frac{11 \chi }{2} & 69 \chi +79435170 \\\hline
\end{array}
\end{align*}
}
	\caption{Genus zero Gopakumar-Vafa invariants of degree zero with respect to the base of the $K3$ fibration for geometries where the intersection form on the polarization lattice is~\eqref{eqn:insnst} for $n=1$.}
	\label{eqn:gv3}
\end{table}

\paragraph{Genus one fibrations with two-sections}
Let us finally comment on the geometries with two-sections that have been discussed in~\cite{Cota:2019cjx}.
Three $K3$ hypersurfaces can be obtained from the polytope with points
\begin{align}
\begin{blockarray}{crrr}
\begin{block}{c(rrr)}
	\nu^1& 1& 0& 0\\
	\nu^2&-1& 1& 0\\
	\nu^3&-1&-1& 0\\
	\nu^4&-2& n& 1\\
	\nu^5& 0& 0&-1\\
\end{block}
\end{blockarray}\,,
\label{eqn:k3points2}
\end{align}
for $n\in\{0,1,2\}$.
The intersection forms on the corresponding Néron-Severi lattice are
\begin{align}
	I^{(2)}_n=\left(\begin{array}{cc}2n&2\\2&0\end{array}\right)\,.
\end{align}
The case $n=2$ corresponds to a $K3$ with Picard rank three but again a part of $\text{NS}_{K3}$ is generated by non-toric divisors that combine into a toric two-section.
Note that the intersection forms $I^{(2)}_0$ and $I^{(2)}_2$ are related by a change of basis.
In~\cite{Banlaki:2018pcc,Cota:2019cjx} the Gopakumar-Vafa invariants of Calabi-Yau threefolds that exhibit a $K3$ fibration and a genus one fibration with two-sections haven been compared against the heterotic one-loop calculation
for compactifications on $(K3\times T^2)/\mathbb{Z}_2$ from~\cite{Chattopadhyaya:2016xpa,Chattopadhyaya:2017zul}.
We note that the matching cases are exactly those where the polarization lattice is $I^{(2)}_0$.
\begin{table}
{\tiny
\begin{align*}
\begin{array}{|c|cccccc|}
	\hline d_B\backslash d_F&0&1&2&3&4&5\\\hline
	0&0 & 168 & 168 & 144 & 168 & 168 \\
	1&54 & 1080 & 9504 & 55080 & 258876 & 1045440 \\
	2& 0 & 1080 & 55080 & 1045440 & 12531888 & 112746384 \\
	3& 0 & 168 & 94248 & 5686200 & 159172380 & 2868991776 \\
	4& 0 & 0 & 55080 & 12531888 & 828397800 & 29153182176 \\
	5& 0 & 0 & 9504 & 12531888 & 2115255492 & 147357745992 \\\hline
\end{array}
\end{align*}
}
	\caption{Genus zero Gopakumar-Vafa invariants of degree zero with respect to the base of the $K3$ fibration for geometries where the intersection form on the polarization lattice is~\eqref{eqn:insnst} for $n=0$.}
	\label{eqn:gv2}
\end{table}

\subsection{All genus checks via the modular bootstrap}
We now want to use the modular bootstrap~\cite{Huang:2015sta,Cota:2019cjx} in order to obtain all-genus expressions for the topological string amplitudes.
Before we apply it to the geometries constructed above, let us briefly review the general technique.

We assume that a Calabi-Yau threefold $M$ is genus one fibered with $N$-sections over a rational base $B$ such that $h^{1,1}(M)=1+h^{1,1}(B)$ and $N\le 4$.
More general cases are covered in~\cite{Cota:2019cjx} but will not be relevant for our discussion.
A basis of divisors is given by the three section $E_0$, as well as a set of vertical divisors $D_i=\pi^{-1}\tilde{D}_i,\,i=1,...,b_2(B)$ for a basis of divisors $\tilde{D}_i$ on $B$.
There is a second set of vertical divisors $D'_i,\,i=1,...,b_2(B)$ that is dual to the curves
\begin{align}
	C_i=\frac{1}{N}E_0\cdot D_i\,.
\end{align}
Finally, one can find a vertical divisor $D$, such that $\tilde{E}_0=E_0+D$ is orthogonal to all of those curves.
The complexified K\"ahler form $\omega$ on $M$ can then be parametrized as
\begin{align}
	\omega=\tau\cdot \tilde{E}_0+{\sum}_i\tilde{t}_i\cdot D'_i\,,
\end{align}
such that $\tau$ is the complexified volume of the generic fiber and $\tilde{t}_i,\,i=1,...,h^{1,1}(B)$ are complexified volumes of the dual curves.
To make the modular structure manifest one needs to introduce shifted K\"ahler parameters
\begin{align}
	t_i=\tilde{t}_i+\frac{\tilde{a}_i}{2N}\tau\,,
\end{align}
where the coefficients $\tilde{a}_i$ are defined as
\begin{align}
	\tilde{a}_i=\int_M\tilde{E}_0^2\cdot D_i\,.
\end{align}

One can now expand the topological string partition function $Z_{\text{top.}}$ on $M$ as
\begin{align}
	Z_{\text{top.}}(\tau,t,\lambda)=Z_0(\tau,\lambda)\left(1+\sum\limits_{\beta\in H_2(B)}Z_\beta(\tau,\lambda)Q^\beta\right)\,,
\end{align}
where $\lambda$ is the topological string coupling constant and $Q^\beta=\exp(2\pi i t\cdot\beta)$.
The coefficients $Z_\beta(\tau,\lambda)$ are conjectured to be meromorphic Jacobi forms of weight $0$ and index $\frac12\beta\cdot(\beta-c_1(B))$ with respect to $\lambda$.
More precisely, they take the form
\begin{align}
	Z_\beta(\tau,\lambda)=\frac{\Delta_{2N}(\tau)^{1-\frac{r_\beta}{N}}}{\eta(N\tau)^{12 \beta\cdot c_1(B)}}\frac{\phi_\beta(\tau,\lambda)}{\prod_{i=1}^{b_2(B)}\prod_{k=1}^{\beta_i}\phi_{-2,1}(N\tau,k\lambda)}\,,
	\label{eqn:bootstrapansatz}
\end{align}
where $\Delta_{2N}(\tau)$ is a particular modular form for $\Gamma_1(N)$, $\eta(\tau)$ is the Dedekind eta function and $\phi_{-2,1}(\tau,\lambda)$ is a weak Jacobi form.
The exponent of $\Delta_{2N}$ is the smallest positive solution to the congruence condition
\begin{align}
	1-\frac{r_\beta}{N}\equiv\frac12\left[Nc_1(B)-\frac{\tilde{a}}{N}\right]\cdot \beta\text{ mod }1\,.
\end{align}
The numerator $\phi_\beta(\tau,\lambda)$ is some weak Jacobi form for $\Gamma_1(N)$ and for low degree $\beta$ it can be fixed, e.g. from the knowledge of some Gopakumar-Vafa invariants.

Let us now apply this procedure to the three-section geometries that we constructed from $\Delta^{(3)}_n$ in the previous section.
The toric data of the four-dimensional ambient space is as follows:
\begin{align*}
\begin{blockarray}{crrrrrrrl}
	&&&&&&&\\
\begin{block}{c(rrrr|rrr)l}
	u& 1& 0& 0& 0& 1& *& *&\leftarrow\text{3-section }E_0\\
	v& 0& 1& 0& 0& 1& *& *&\leftarrow\text{3-section }E_0-n\cdot D_1+(\nu_2-\nu_1)\cdot D_2\\
	w&-1&-1& 0& 0& 1& *& *&\leftarrow\text{3-section }E_0-n\cdot D_1-\nu_1\cdot D_2\\
	a_1& n& 0& 1& 0& 0& 1& 0&\leftarrow\text{vertical divisor }D_2'=\pi^{-1}(B-\nu_3F)\\
	a_2& 0& 0&-1& 0& 0& 1&\nu_3&\leftarrow\text{vertical divisor }D_1=\pi^{-1}(B)\\
	b_1& \nu_1&\nu_2&\nu_3& 1& 0& 0& 1&\leftarrow\text{vertical divisor }D_2=D_1'=\pi^{-1}(F)\\
	b_2& 0& 0& 0&-1& 0& 0& 1&\phantom{x}\hspace{1.5cm}\text{\ditto}\\
	   & 0& 0& 0& 0&-3&-1& 0&\\
\end{block}
\end{blockarray}
\label{eqn:pqpointsExample}
\end{align*}
Recall that we used lattice automorphisms to fix $\nu_3\le 0$ such that $B$ and $F$ are the base and fiber of the Hirzebruch surface $\mathbb{F}_{-\nu_3}$ that is the base of the genus one fibration.
While one can give closed expressions for the empty values in the Mori-cone, they are somewhat bulky and we do not require them in the following discussion.

We choose $E_0=[u]$ as the ``zero'' three-section and find
\begin{align}
	E_0^2\cdot D_1=-4\cdot \nu_1+2\cdot \nu_2+\nu_3+2\,,\quad E_0^2\cdot D_2=2-4\cdot n\,.
\end{align}
as intersections on the generic Calabi-Yau hypersurface.
This implies that
\begin{align}
	\tilde{E}_0=E_0+\frac23(2\cdot n-1)D_1+\frac13(4\cdot n+4\cdot\nu_1-2\cdot \nu_2-\nu_3-4)D_2\,,
\end{align}
is orthogonal to the curves $E_0\cdot D_{1/2}$ and we calculate
\begin{align}
	\tilde{a}_1=\tilde{E}_0^2\cdot D_2=4\cdot n-2\,,\quad \tilde{a}_2=\tilde{E}^2\cdot D_1=4\cdot\nu_1-2\cdot\nu_2-\nu_3-2\,.
\end{align}
The correct parametrization of the K\"ahler form to obtain modular amplitudes is then
\begin{align}
	\omega=\tau\cdot\tilde{E}_0+\left(t_1-\frac16\tilde{a}_1\right)D_1'+\left(t_2-\frac16\tilde{a}_2\right)D_2'\,.
\end{align}

We want to apply the modular bootstrap to obtain all genus amplitudes with respect to the $K3$ fiber, i.e. we are interested in $Z_\beta$ for $\beta=k\cdot F$.
The exponent $r_\beta$ in~\eqref{eqn:bootstrapansatz} is then given by
\begin{align}
	1-\frac{r_{k\cdot F}}{3}\equiv\frac{k}{3}(2\cdot n-1)\text{ mod }1\,.
\end{align}
For $n\in\{-1,2\}$ and $k=1$ we use genus zero Gromov-Witten invariants to fix the ansatz
\begin{align}
	Z_1(\tau,\lambda)=\frac{1}{48}\frac{\Delta_6(\tau)}{\eta(3\tau)^{24}\phi_{-2,1}(3\tau,\lambda)}\left[(120+\chi)E_4(\tau)-9(152+\chi)\mathcal{E}_3(\tau)^2\right]\,,
	\label{eqn:zex}
\end{align}
where we refer to the Appendix~\ref{app:conventions} for the definition of the modular and Jacobi forms.
From this we can extract Gopakumar-Vafa invariants for arbitrary fiber degrees and genera and the results for low orders are listed in table~\ref{tab:bootstrapresults}.
\begin{table}
\begin{align*}
\begin{array}{|c|cccccc|}
  \hline d_F\backslash g& 0 & 1 & 2 & 3 & 4 & 5 \\\hline
 0 & \frac{\chi }{6}+26 & 0 & 0 & 0 & 0 & 0 \\
 1 & \frac{\chi }{2}+240 & 0 & 0 & 0 & 0 & 0 \\
 2 & 1962-3 \chi  & 0 & 0 & 0 & 0 & 0 \\
 3 & \frac{10 \chi }{3}+18016 & -\frac{\chi }{3}-52 & 0 & 0 & 0 & 0 \\
 4 & \frac{15 \chi }{2}+95454 & -\chi -480 & 0 & 0 & 0 & 0 \\
 5 & 413280-30 \chi  & 6 \chi -3924 & 0 & 0 & 0 & 0 \\
 6 & \frac{88 \chi }{3}+1627330 & -\frac{23 \chi }{3}-36188 & \frac{\chi }{2}+78 & 0 & 0 & 0 \\
 7 & 54 \chi +5694624 & -18 \chi -192348 & \frac{3 \chi }{2}+720 & 0 & 0 & 0 \\
 8 & 18353988-195 \chi  & 78 \chi -838332 & 5886-9 \chi  & 0 & 0 & 0 \\
 9 & 170 \chi +55646304 & -80 \chi -3362964 & \frac{38 \chi }{3}+54464 & -\frac{2 \chi }{3}-104 & 0 & 0 \\
 10 & 291 \chi +159217686 & -157 \chi -11963892 & \frac{61 \chi }{2}+290202 & -2 \chi -960 & 0 & 0 \\\hline
\end{array}
\end{align*}
	\caption{Gopakumar-Vafa invariants of degree zero with respect to the base of the $K3$ fibration and degree one with respect to the base of the $K3$ fiber.}
	\label{tab:bootstrapresults}
\end{table}
This matches with the predictions from the heterotic calculation and provides another strong check of the duality.
 
 \section{Conclusions}
In this paper we have explicitly calculated the new supersymmetric index for several heterotic compactifications on $(K3\times T^2)/\mathbb{Z}_3$.
To this end we considered a $T^4/\mathbb{Z}_3$ orbifold limit of the $K3$ with all of the five possible embeddings of the $\mathbb{Z}_3$ action into the gauge group.
The index encodes the gauge and gravitational coupling of the ${\cal N}=2$ effective theory that arise from heterotic one-loop amplitudes 
and leads to predictions for certain Gopakumar-Vafa invariants of dual Calabi-Yau compactification spaces.
This generalizes earlier calculations for compactifications on $K3\times T^2$~\cite{Harvey:1995fq,Marino:1998pg,Stieberger:1998yi} as well as
the subsequent extensions to $(K3\times T^2)/\mathbb{Z}_N$, where for $N=3$ only the standard embedding has been considered~\cite{Chattopadhyaya:2016xpa,Chattopadhyaya:2017zul}.
In particular, we obtain two vacua where the hypermultiplet spectrum enables us to maximally higgs the gauge group to $U(1)^3$.

Let us note that for all choices of embeddings except for the standard embedding, the new supersymmetric index depends only on the difference $N_h-N_v$ of the number of vector- and hypermultiplets. 
This is the same behaviour that was observed for \textit{all} embeddings when the heterotic string is compactified on $K3\times T^2$~\cite{Stieberger:1998yi} or $(K3\times T^2)/\mathbb{Z}_2$~\cite{Chattopadhyaya:2016xpa}.
It would be interesting to understand the reason why the standard embedding for $\mathbb{Z}_3$ orbifolds is special. 
We hope to come back to this question in future work.

We then systematically constructed candidate Calabi-Yau duals that exhibit both a $K3$ and a genus one fibration structure,
as well as three sections with respect to the latter.
In this way we could show that the enumerative invariants exhibit the correct structure if and only if the intersection form on the
polarization lattice is three times the intersection form on the Narain lattice $\Gamma^{1,1}$.
Physically, this can be interpreted as the requirement that the dual heterotic compactification exhibits T-duality.
For each of the two heterotic models that can be higgsed to $U(1)^3$ we find Calabi-Yau manifolds that reproduce the predicted enumerative invariants.
We apply the modular bootstrap~\cite{Cota:2019cjx} to obtain all genus results for a subset of the topological string amplitudes and in that way provide a strong check of the duality.
Our discussion generalizes and provides a more systematic understanding of earlier constructions of Calabi-Yau duals for heterotic compactifications on $(K3\times T^2)/\mathbb{Z}_2$~\cite{Kachru:1997bz,Banlaki:2018pcc,Cota:2019cjx}.

Multiple geometries that match the same new supersymmetric index are distinguished by their classical intersection numbers and by those enumerative invariants that lead to predictions for what from the heterotic perspective are non-pertubative corrections.
This situation is analogous to that encountered already for heterotic compactifications on $K3\times T^2$.
It is well known that the embeddings of the gauge connection with instanton numbers $(12-n,12+n)$ for $n=0,1,2$ produce the same new supersymmetric index
and are respectively dual to type IIA compactifications on generic elliptic fibrations over the Hirzebruch surfaces $\mathbb{F}_n$~\cite{Morrison:1996pp}.
However, it is quite difficult to derive the heterotic instanton numbers from the geometry of the Calabi-Yau or the classical intersection numbers of the Calabi-Yau from a heterotic calculation, see e.g.~\cite{Braun:2016sks} for a discussion of this issue.
It will be very interesting to study this problem in the context of heterotic compactifications on $(K3\times T^2)/\mathbb{Z}_N$ and their dual Calabi-Yau manifolds.
We leave this as another question that we hope to address in future work.

\appendix
\section{Conventions for modular forms}
\label{app:conventions}
In this Appendix we summarize the definitions of the various modular and Jacobi forms that appear throughout the paper.
The Dedekind eta function is defined as
\begin{align}
	\eta(\tau)=q^{\frac{1}{24}}\prod\limits_{n=1}^\infty(1-q^n)\,,
\end{align}
where $q=\exp(2\pi i\tau)$. 
It is a modular form of weight $1/2$ with a multiplier system and transforms as
\begin{align}
	\eta(\tau+1)=e^{\frac{\pi i}{12}}\eta(\tau)\,,\quad\eta(-1/\tau)=\sqrt{-i\tau}\eta(\tau)\,.
\end{align}
The classical Eisenstein series can be written as
\begin{align}
	E_4(\tau)=1+240\sum\limits_{n=1}^\infty\frac{n^3q^n}{1-q^n}\,,\quad E_6(\tau)=1-504\sum\limits_{n=1}^\infty\frac{n^5q^n}{1-q^n}\,,
\end{align}
and are modular forms of respective weight four and six.

We will also need to consider modular forms for the congruence subgroup $\Gamma_1(3)$ of the modular group.
The corresponding ring can be generated by three functions $E_4,E_6$ and $\mathcal{E}_3$, where $\mathcal{E}_3$ is a holomorphic modular form of weight two that can be expressed
in terms of the Dedekind eta function via
\begin{align}
	\mathcal{E}_3(\tau)=-\frac{24}{2\pi i}\partial_\tau\log\left(\frac{\eta(\tau)}{\eta(3\tau)}\right)\,.
\end{align}
Moreover, the unique weight six modular form for $\Gamma_1(3)$ with a second order zero at $\tau\rightarrow i\infty$ is given by
\begin{align}
\Delta_6(\tau)=\frac{\eta(3\tau)^{18}}{\eta(\tau)^6}\,.
\end{align}

The generalized Jacobi theta function can be written as
\begin{align}
	\theta\left[\begin{array}{c}a\\b\end{array}\right](\tau,z)=\sum\limits_{k\in\mathbb{Z}}q^{\frac12\left(k+\frac{a}{2}\right)^2}e^{\pi i\left(k+\frac{a}{2}\right)b}e^{(2\pi i z)\left(k+\frac{a}{2}\right)}\,,
\end{align}
and one introduces
\begin{align}
	\begin{split}
	\theta_1(\tau,z)=&\theta\left[\begin{array}{c}1\\1\end{array}\right](\tau,z)\,,\quad \theta_2(\tau,z)=\theta\left[\begin{array}{c}1\\0\end{array}\right](\tau,z)\,,\\
	\theta_3(\tau,z)=&\theta\left[\begin{array}{c}0\\0\end{array}\right](\tau,z)\,,\quad \theta_4(\tau,z)=\theta\left[\begin{array}{c}0\\1\end{array}\right](\tau,z)\,.
	\end{split}
\end{align}
This determines also the weak Jacobi form $\phi_{-2,1}(\tau,z)$ of weight $-2$ and index $1$ via
\begin{align}
	\phi_{-2,1}(\tau,z)=-\frac{\theta_1(\tau,z)^2}{\eta(\tau)^6}\,,
\end{align}
that we need for the modular bootstrap.

\section{Higgsing of the gauge group}
\label{app:higgsing}
In this Appendix we will explain in some detail how the gauge group of the  models of table \ref{3anh} can be higgsed by giving vacuum expectation values (vev's) to scalars in the hypermultiplets. We take the necessary branching rules from \cite{Yamatsu:2015npn}.  For more on higgsing see for example~\cite{Srednicki:1019751, Slansky:1981yr}.

For concreteness sake we will study the third model of table~\ref{3anh}. 
Before  higgsing we have the following  gauge group and matter content, coming from the twisted and untwisted sector
\begin{eqnarray}
&& SO(14)\times SO(14) \times U(1)^2\\\nonumber
&& 4\bold{(14,1)}+4\bold{(1,14)+(64,1)+(1,64)}+8\bold{(1,1)}\,.
\end{eqnarray}
The matter content is labelled by the representations under the two $SO(14)$ groups, i.e. we have left out the $U(1)$ charges.
Counting degrees of freedom we find the numbers of vector- and hypermultiplets $N_v=184$ and $N_h=248$.
We will start by explaining how to higgs the first $SO(14)$ factor. 

For this we notice the following branching rules 
\begin{eqnarray}
SO(14) \supset SO(13):\: \bold{14 \rightarrow 13 +1} \:,\:
\bold{64 \rightarrow 64}.
\end{eqnarray}
Giving a vev to one  $\bold {14}$ will break  $SO(14)$ to $SO(13)$.
One $\bold {13}$ will get `eaten' by the broken generators of the gauge fields (turning them massive)  and the result is the following gauge group and matter spectrum
\begin{eqnarray}
&& SO(13)\times SO(14) \times U(1)^2 \:,\:\\ \nonumber
&&  3\bold{(13,1)}+4\bold{(1,14)+(64,1)+(1,64)}+12\bold{(1,1)}.
\end{eqnarray}
We may check that $N_h-N_v=235-171=64$ is unchanged.
In the next step we  give a vev to  a $\bold {13}$ and use the branching rules
\begin{eqnarray}
SO(13) \supset SO(12):\: \bold{13 \rightarrow 12 +1} \:,\:
\bold{64 \rightarrow 32 + 32^\prime}.
\end{eqnarray}
Similar to before a $\bold {12}$  gets `eaten' by the broken generators of $SO(13)$ and we find the  following gauge group and matter spectrum after higgsing
\begin{eqnarray}
&& SO(12)\times SO(14) \times U(1)^2 \:,\:\\ \nonumber
&& 2\bold{(12,1)}+4\bold{(1,14)+(32,1)+ (32^\prime,1)+(1,64)}+15\bold{(1,1)}.
\end{eqnarray}
Two more similar steps lead to the following gauge group and matter spectrum 
\begin{eqnarray}
&& SO(10)\times SO(14) \times U(1)^2 \:,\:\\ \nonumber
&& 4\bold{(1,14)}+2\bold{(16,1)}+2(\bold{\overline{16},1)+(1,64)}+18\bold{(1,1)}.
\end{eqnarray}
The branching rules
\begin{eqnarray}
SO(10) \supset SU(5) \times U(1): \bold{16 \rightarrow 10(1) + \bar 5(-3) + 1(5)},\\ \nonumber \bold{ \overline{16} \rightarrow \bar 10(-1) + 5(3) + 1(-5)}
\end{eqnarray}
(the number in brackets give the $U(1)$ charge) indicate that  we can give a vev to $\bold{16}$ and $\bold{\overline{16}}$ thereby breaking $SO(10)$ to $SU(5)$  where $\bold{10,\overline{10}}$ and one scalar will get `eaten' by the broken generators. The gauge group and spectrum after this step of  higgsing are thus
\begin{eqnarray}
&& SU(5)\times SO(14) \times U(1)^2 \:,\:\\ \nonumber
&& 4\bold{(1,14)}+2\bold{(5,1)}+2(\bold{\overline{5},1)+ (10,1)+(\overline{10},1) + (1,64)}+ 21 \bold{(1,1)}.
\end{eqnarray}
In the next step we give a vev to $\bold{5,\bar 5}$ and use the branching rules
\begin{eqnarray}
SU(5)\supset SU(4)\times U(1):\: \bold{5\rightarrow 4 (1)+ 1(-4) },\\\nonumber
\bold{10 \rightarrow 4 (-3) + 6 (2)} \end{eqnarray}
to obtain the gauge group and spectrum
\begin{eqnarray}
&&SU(4)\times SO(14) \times U(1)^2 \:,\:\\ \nonumber
&&4\bold{(1,14)}+2\bold{(4,1)}+2(\bold{\overline{4},1)}+ \bold{(6,1)}+ \bold{(\bar 6 ,1) + (1,64)}+ 24 \bold{(1,1)}.
\end{eqnarray}
We proceed by breaking $SU(4)$ to $SU(3)$ by giving a vev to $\bold{4,\bar 4}$ and using the branching rules
\begin{eqnarray}
SU(4) \supset SU(3)\times U(1): \bold{4 \rightarrow 3(1) + 1 (-3)} , \\ \nonumber 
\bold{6 \rightarrow 3 (-2) + \bar 3 (2)}.
\end{eqnarray}
We obtain the following gauge group and spectrum
\begin{eqnarray}
&& SU(3)\times SO(14) \times U(1)^2 \:,\:\\ \nonumber
&& 4 \bold{(1,14)} +3\bold{(3,1)} + 3 \bold{(\overline{3},1) + (1,64)} + 27 \bold{(1,1)}.
\end{eqnarray}
We can continue in similar manner using $3\bold{(3,1)}$ and  $3\bold{(\bar 3,1)}$ to higgs $SU(3)$ completely and end up with
\begin{eqnarray}
&& SO(14) \times U(1)^2 \:,\:\\ \nonumber
&& 4 \bold{(1,14) + (1,64)} + 37 \bold{(1,1)}.
\end{eqnarray}
In the same way we may higgs the second $SO(14)$. The two $U(1)$ factors in the gauge group may be higgsed by any of the charged scalars. Thus we end up with a completely higgsed gauge group and 64 neutral scalars.

\bibliography{ref-col}
\bibliographystyle{JHEP} 

\end{document}